\documentclass[twocolumn,tighten]{aastex631}   
\usepackage{amsmath}

\newcommand{\ee}[1]{\mbox{${} \times 10^{#1}$}}    
\newcommand{\eten}[1]{\mbox{$10^{#1}$}}            
\newcommand{\msun}{\mbox{M$_\odot$}}               

\newcommand{\co}{\textsuperscript{13}CO}
\newcommand{\coo}{\mbox{$^{13}$CO}}
\newcommand{\cotw}{\mbox{$^{12}$CO}}

\newcommand{\kms}{km s$^{-1}$}
\newcommand{\kkms}{K km s$^{-1}$}

\newcommand{\hh}{\mbox{{\rm H}$_2$}}
\newcommand{\hcop}{\mbox{{\rm HCO}$^+$}}

\newcommand{\hcn}{\mbox{{\rm HCN}}}

\newcommand{\hii}{\mbox{\ion{H}{2}}}
\newcommand{\jj}[2]{\mbox{$J = #1\rightarrow#2$}}
\newcommand{\gal}{\mbox{Galactocentric }}
\newcommand{\rg}{\mbox{$R_{G}$}}

\newcommand{\mcloud}{\mbox{$M_{\rm cloud}$}}

\newcommand{\mdense}{\mbox{$M_{\rm dense}$}}

\newcommand{\av}{\mbox{$A_{\rm V}$}}
\newcommand\cmv{\mbox{cm$^{-3}$}}

\newcommand{\lin}{\mbox{$L_{\rm in}$}}             
\newcommand{\ltot}{\mbox{$L_{\rm tot}$}}           
\newcommand{\ain}{\mbox{$\alpha_{\rm in}$}}        
\newcommand{\at}{\mbox{$\alpha_{\rm tot}$}}        
\newcommand{\fl}{\mbox{$f_{\rm L}$}}
\newcommand{\nh}{\mbox{$\rm N(H_{2})$}}
\newcommand{\clouda}{\mbox{S206}}
\newcommand{\cloudb}{\mbox{S208}}
\newcommand{\cloudc}{\mbox{S212}}
\newcommand{\cloudd}{\mbox{S228}}
\newcommand{\cloude}{\mbox{S235}}
\newcommand{\cloudf}{\mbox{S252}}
\newcommand{\cloudg}{\mbox{S254}}

\shorttitle {Outer Galaxy}    
\shortauthors{Patra et al.}   
\graphicspath{{./}{figures/}}

\begin{document}

\title{Tracers of Dense Gas in the Outer Galaxy}   

\correspondingauthor{Sudeshna Patra}
\email{inspire.sudeshna@gmail.com}
\author[0000-0002-3577-6488]{Sudeshna Patra}
\affiliation{Indian Institute of Science Education and Research (IISER) Tirupati,
Rami Reddy Nagar, Karakambadi Road, Mangalam(P.O.), Tirupati 517507, India}

\author[0000-0001-5175-1777]{Neal J. Evans II}
\affiliation{Department of Astronomy The University of Texas at Austin,
2515 Speedway, Stop C1400 Austin, Texas 78712-1205, USA}
\affiliation{Korea Astronomy and Space Science Institute,
776 Daedeokdae-ro, Yuseong-gu Daejeon, 34055, Korea}
\affiliation{Humanitas College, Global Campus, Kyung Hee University,
Yongin-shi 17104, Korea}

\author[0000-0003-2412-7092]{Kee-Tae Kim}
\affiliation{Korea Astronomy and Space Science Institute,
776 Daedeokdae-ro, Yuseong-gu Daejeon, 34055, Korea}
\affiliation{University of Science and Technology, Korea (UST),
217 Gajeong-ro, Yuseong-gu, Daejeon 34113, Republic of Korea}

\author[0000-0002-3871-010X]{Mark Heyer}
\affiliation{Department of Astronomy University of Massachusetts Amherst,
Massachusetts 01003, USA}

\author[0000-0002-5094-6393]{Jens Kauffmann}
\affiliation{Haystack Observatory MIT,
99 Milstone Rd. Westford, MA 01886 USA}

\author[0000-0003-4908-4404]{Jessy Jose}
\affiliation{Indian Institute of Science Education and Research (IISER) Tirupati,
Rami Reddy Nagar, Karakambadi Road, Mangalam(P.O.), Tirupati 517507, India}

\author[0000-0002-9431-6297]{Manash R. Samal}
\affiliation{Astronomy \& Astrophysics Division, Physical Research Laboratory,
Ahmedabad, 380009, State of Gujarat, India}

\author[0000-0001-7151-0882]{Swagat R. Das}
\affiliation{Indian Institute of Science Education and Research (IISER) Tirupati,
Rami Reddy Nagar, Karakambadi Road, Mangalam(P.O.), Tirupati 517507, India}

\begin{abstract}
We  have mapped \hcn\ and \hcop\ (\jj10) line emission toward a sample of seven star-forming regions (with $\mathrm{12+log[O/H]}$ range from 8.34 to 8.69) in the outer Milky Way (Galactocentric distance $> 9.5$ kpc),  using the 14-meter radio telescope of the Taeduk Radio Astronomy Observatory (TRAO). 
We compare these two molecular lines with  other conventional tracers of dense gas, millimeter-wave continuum emission from dust and extinction thresholds ($\av \ge 8$ mag), inferred from the \coo\ line data. 
\hcn\ and \hcop\ correlate better with the millimeter emission than with the extinction criterion.
A significant amount of luminosity comes from regions below the extinction criterion and outside the millimeter clump for all the clouds. 
The average fraction of \hcn\ luminosity from within the regions with $\av \ge 8$ mag is $0.343 \pm 0.225$; for the regions of millimeter emission, it is $0.478 \pm 0.149$.  
Based on a comparison with column density maps from Herschel, \hcn\ and \hcop\ trace dense gas in high column density regions better than does \coo. 
\hcop\ is less concentrated than \hcn\ for outer Galaxy targets, in contrast with the inner Galaxy sample, suggesting that metallicity may affect the interpretation of tracers of dense gas.
The conversion factor between the dense gas mass (\mdense) and line luminosities of \hcn\ and \hcop, when integrated over the whole cloud, is comparable with factors used in extragalactic studies.
\end{abstract}


\section{Introduction} \label{sec:intro}

Detailed observation of molecular clouds in the Milky Way \citep{2010ApJ...723.1019H, 2010ApJ...724..687L, 2012ApJ...745..190L,2014ApJ...782..114E,2016ApJ...831...73V} and studies of external galaxies \citep{2004ApJ...606..271G} have shown that star formation is more accurately predicted by the amount of dense molecular gas rather than by the total amount of molecular gas. The molecules with high dipole moment and high critical density, such as \hcn, \hcop\ are used as dense gas tracers. The pioneering studies by \citet{2004ApJ...606..271G,2004ApJS..152...63G} revealed a tight linear correlation between the luminosity of the \hcn\ \jj10 transition (as a tracer of the mass of dense molecular gas) and the far-infrared luminosity, indicative of the star formation rate (SFR) for whole galaxies. 
Later works  extended this relation to resolved  structures  in other galaxies \citep{2019ApJ...880..127J, 2022ApJ...930..170H} or even to massive clumps in the Milky Way \citep{2005ApJ...635L.173W,2016ApJ...829...59L,2016ApJ...824...29S, 2017A&A...604A..74S}.

Recent studies \citep{2017A&A...599A..98P, 2017A&A...605L...5K,2020ApJ...894..103E,2020ApJ...891...66N} have questioned how well the lowest $J$ transitions of \hcn\ and \hcop\ trace the dense gas. These authors report that \hcn\ and \hcop \jj10 emission is easily detectable from the diffuse part of the cloud with typical density $\mathrm{100-500\ cm^{-3}}$. The concept of detecting a particular line above or around the critical density ($n_{c}$) is oversimplified \citep{1989RMxAA..18...21E}. 
In reality, multilevel excitation effects and radiative trapping effects all tend to lower the effective excitation density ($n_{eff}$), defined as the density needed to detect a molecular line with integrated intensity of 1 \kkms\ \citep{1999ARA&A..37..311E,2015PASP..127..299S}.  
For the low-$J$ transitions the effective excitation densities are typically 1-2 orders of magnitude lower than the critical densities.  The values of $n_{eff}$ for \hcn\ and \hcop\ are $4.5\times 10^{3}\ \mathrm{cm^{-3}}$ and $5.3 \times 10^{2}\ \mathrm{cm^{-3}}$ at 20 K, respectively \citep{2015PASP..127..299S}. For modern surveys with low RMS noise, lines at the detection limits can be produced by gas with density as low as 50-100 \cmv, explaining the emission from extended translucent regions where the density of gas is much lower than the critical densities of \hcn, \hcop\ \citep{2020ApJ...894..103E}. 

\citet{2017A&A...599A..98P} found that only a small fraction of the luminosity of \hcn (18\%) and \hcop (16\%) arises from regions with $\av > 15$ mag for Orion B. \citet{2017A&A...605L...5K} found that \hcn (\jj10) mainly traces gas with $\av  \approx 6$ mag, or $n \approx 870\ \mathrm{cm^{-3}}$ towards Orion A. A substantial fraction (44-100\%) of the total line luminosities of \hcn\ and \hcop arises outside the $\av > 8$ mag region often associated with star formation for 6 inner Galaxy clouds \citep{2020ApJ...894..103E}.
Most dense cores and YSOs are found at $\av \geq 8$ mag 
\citep{2010ApJ...723.1019H, 2010ApJ...724..687L, 2012ApJ...745..190L}.
 
None of the studies described above considered the star forming regions in the outer Galaxy, i.e., clusters having \gal\ distance (\rg) greater than $\sim 8.2$ kpc.  The star-forming regions in the outer Galaxy (low-density regime) behave distinctively from the environment around the solar neighborhood (intermediate density regime) and also from the Central Molecular Zone (CMZ, high-density regime) \citep{2012ARA&A..50..531K}. The physical and chemical conditions in star-forming regions are dependent on the environment, which is in turn a function of Galactocentric radius ($R_{G}$). 
 
The Milky Way and other normal spiral galaxies are known to have a negative gradient of metallicity with \rg\ \citep{1971ApJ...168..327S}. Many studies have found a radial abundance gradient of oxygen, nitrogen, etc. in the Milky Way \citep{2000MNRAS.311..329D,2017MNRAS.471..987E,2018MNRAS.478.2315E}. Most recently, \citet{2022MNRAS.tmp...13M} report linear radial abundance gradients for $\mathrm{O/H}$, $\mathrm{N/H}$  with slopes of $-0.044 \pm 0.009 \ \rm dex \ kpc^{-1}$, $-0.063 \pm 0.009 \ \rm dex \ kpc^{-1}$,  respectively, for $R_G =$ 4.88 to 17 kpc and for $\mathrm{C/H}$ the slope is $-0.077 \pm 0.019 \ \rm dex \ kpc^{-1}$ for $R_G = 6.46$ to $11.6$ kpc.
The efficiency of gas cooling and dust shielding processes decrease at lower metallicity, thereby affecting the condensation and fragmentation of giant molecular clouds (GMCs). Conversely, the lower interstellar radiation field and cosmic-ray fluxes at larger $R_{G}$ decreases  the gas heating \citep{2011ApJ...738...27B}. The balance between these processes sets the gas and dust physical conditions may cause different star formation outcomes from those in the CMZ or Solar neighborhood.
While studies of nearby galaxies (LMC, SMC) with low metallicity are used to understand these effects \citep{2020A&A...643A..63G}, the outer Milky Way is much closer to us. The outer Galaxy targets provide a less confusing view of the ISM as they are widely separated and there is no blending of emission. This clear perspective of the outer Galaxy enables studies of individual giant molecular clouds from which global properties of star formation can be estimated. 
Thus, studying the star forming regions in the outer Galaxy can bridge the gap between the star formation studies in Galactic and extragalactic environments.
 
To inform interpretation of extragalactic studies, we investigate the distribution of \hcn\ and \hcop\ \jj10 emission over the full area of molecular clouds and range of physical conditions. This work complements the inner Galaxy work \citep{2020ApJ...894..103E} for the outer Galaxy. 



\section{Sample} \label{sec:Sample}

 We selected 7 targets beyond the solar circle with $\rg > 9.5 \ \mbox{kpc}$ with diversity in properties such as cloud mass, physical size, number of massive stars, metallicity etc (see Table \ref{tab:cloud details}). We also considered the availability of molecular data (\cotw \ and \coo \ emission from the Extended Outer Galaxy survey, see \ref{subsec:ancillary_data}), dust emission from the BGPS survey \citep{2013ApJS..208...14G} and FIR emission from Herschel data \citep{2017MNRAS.471.2730M}. 
 Several distance values are available in the literature based on kinematic or stellar measurements. In order to maintain uniformity we adopt the GAIA EDR3 based distance values of all the targets (except \cloudf) from \cite{2022MNRAS.tmp...13M} and use the same methodology to obtain the distance for \cloudf.
 The distance of the Sun from the Galactic Center is taken as 8.2 kpc \citep{2019A&A...625L..10G}. \rg\ values of our sample range between 9.85 to 14.76 kpc (see Table \ref{tab:cloud details}). 

 The spectral type(s) (SpT) of the main ionizing star(s) and \mbox{12+log[O/H]} values of each target and their respective references  are also given in Table \ref{tab:cloud details}.
 We mapped the \hcn\ and \hcop\ (\jj10) emission from the parent clouds of 7 \hii\ regions over an area based on \coo\ emission (see details in Section \ref{sec:Obs}). 
 The cloud mass (\mcloud) is obtained using the CO data described in \S \ref{sec:Obs} and $\mcloud=\alpha_{\mathrm{CO}}L_{\mathrm{CO}}$ with $\alpha_{\mathrm{CO}}= 4.3\ \mathrm{M_{\odot}\ (K\ km\ s^{-1} pc^{2})^{-1}}$, where $L_{\mathrm{CO}}$ is the luminosity expressed in $\mathrm{K\ km\ s^{-1} pc^{2}}$ \citep{2013ARA&A..51..207B}.
 A brief description of each region can be found in \S Appendix \ref{app:target}. 

\section{Observations and Data Sets} \label{sec:Obs}

\subsection{\hcn\ and \hcop\ Observations with TRAO }
 We mapped the seven outer Galaxy clouds in the HCN (\jj10, 88.631847 GHz) and \hcop (\jj10, 89.1885296 GHz) lines simultaneously at Taeduk Radio Astronomy Observatory (TRAO) in January 2020. The TRAO telescope is equipped with the SEcond QUabbin Optical Image Array (SEQUOIA-TRAO), a multi-beam receiver with 16 pixels in a $4 \times 4$ array and operates in 85-115 GHz range. The back-end has a 2G FFT (Fast Fourier Transform) Spectrometer. The instrumental properties are described in \citet{2019JKAS...52..227J}. The TRAO has a main-beam size of 58\arcsec\ at these frequencies and a main-beam efficiency (defined so that $T_{\rm mb} = T_A^*/(f_b \eta_{\rm mb})$, where $T_{\rm mb}$ is the Rayleigh-Jeans main-beam temperature and $f_b$ is the fraction of the beam filled by the source) of 46\% at 89 GHz \citep{2019JKAS...52..227J}. 
 The individual SEQUOIA beams vary in beam size (efficiency) by only a few arcseconds (few percent). The beams are very circular and beam efficiencies vary less than $3\%$ with elevation angle. 

We used the OTF (On-The-Fly) mode to map the clouds. The mapped areas for the individual targets were determined from maps of their $^{13}$CO (\jj10) line emission.  We used a threshold value of 5 times the RMS noise in the \coo\ map to estimate the mapping area for \hcn\ and \hcop\  from the TRAO 14-m telescope. 
The \av\ values corresponding to the threshold values are listed in Table \ref{tab:cloud details}.
All the targets were mapped in Galactic co-ordinates. The mapped areas for individual targets are listed in Table \ref{tab:cloud details}.

The system temperature was $\sim 190 \rm K$ during \hcn\ and \hcop\ observations.  We used $\rm SiO$ maser sources at 86 GHz for pointing and focusing the telescope  every 3 hours. We smoothed the data with the boxcar function to a velocity resolution of about 0.2 \kms, resulting in an RMS sensitivity of 0.06-0.08 K on the $T_{A}^{*}$ scale for all the targets.
We obtained $\sim 150\ \rm hrs$ of observations in total.

The following steps were done to reduce the data.  
We used the OTFTOOL to read the raw on-the-fly data for each map of each tile and then converted them into the CLASS format map after subtracting a first order baseline.
We applied noise weighting and chose the output cell size of $20\arcsec$ in OTFTOOL. 
We did not apply any convolution function to make the maps, so the effective beam size is same as the main-beam size.
Further reduction was done in CLASS format of GILDAS\footnote{\url{https://www.iram.fr/IRAMFR/GILDAS/}}, where we chose the velocity range ($v_{sp}$) to get good baselines on each side of the velocity range of significant emission ($v_{win}$). The reduction details are listed in \S Appendix \ref{app:reduction} Table \ref{tab:reduction}. 
We subtracted a second-order polynomial baseline and converted those files to FITS format using GREG.

\subsection{Ancillary data available} \label{subsec:ancillary_data}
For this work, we have also used the following data sets :
\begin{enumerate}
    \item \textbf{\cotw\ and \co\ (\jj10) Observations:} \cotw\ and \coo\ data for all the targets are extracted 
    from the Exeter-FCRAO, Extended Outer Galaxy survey using the Five College Radio Astronomy Observatory (FCRAO) 14-m telescope. The EXFC survey covered \cotw\ and \coo\ \jj10 emission within two longitude ranges: $55^{\circ}<l<100^{\circ}$ with the Galactic Latitude range $-1.4^{\circ} \le b \le +1.9^{\circ}$ and $135^{\circ}<l<195^{\circ}$ with Galactic Latitude range $-3.6^{\circ}<b<5.6^{\circ}$ \citep{2016ApJ...818..144R}. Our outer Galaxy targets are covered by the latter survey. The angular resolution of \cotw\ and \coo\ are $45\arcsec$ and $48\arcsec$ respectively and the grid size is $22.5\arcsec$. The spectral resolution is 0.127 \kms.
    All of the data have been de-convolved to remove contributions from the antenna error beam and therefore, are in main beam temperature units.
    
    \item \textbf{Millimeter Dust Continuum Emission:} We use 1.1 mm continuum emission data from the Bolocam Galactic Plane Survey (BGPS)\footnote{\citet{https://doi.org/10.26131/irsa482}} using Bolocam on the Caltech Submillimeter Observatory. The BGPS Survey does not cover the outer Galaxy regions uniformly; it covers four targeted regions (IC1396, a region towards the Perseus arm, W3/4/5 and Gem OB1).
    The effective resolution of this survey is $33\arcsec$. We use the mask files of the sources from V2.1 table\footnote{\url{https://irsa.ipac.caltech.edu/data/BOLOCAM_GPS/}}, where the pixel values in the mask file are set with the catalog number where the source is above the $2\sigma$ threshold; otherwise the pixel values are zero \citep{2013ApJS..208...14G}.
    The Bolocat V2.1 contains the 1.1 mm photometric flux of 8594 sources for the aperture radii of $20\arcsec$, $40\arcsec$, and $60\arcsec$ and it also provides flux densities integrated over the source area.
    As this is not a contiguous survey towards the outer Galaxy, not all the targets have BGPS data. The data availability of the targets are indicated by the footnotes in Table \ref{tab:cloud details}.

    \item \textbf{Far-Infrared data:} 
    We have used the column density maps determined from Herschel data \citep{2017MNRAS.471.2730M} and available at the Herschel infrared Galactic Plane (Hi-GAL) survey site table\footnote{\url{http://www.astro.cardiff.ac.uk/research/ViaLactea/}}. These column density maps are derived using the PPMAP procedure \citep{2015MNRAS.454.4282M} on the continuum data in the wavelength range 70-500 $\rm \mu$m from the Hi-GAL survey. The spatial resolution of the maps is $12\arcsec$. Not all the targets in our list are covered by the Hi-GAL survey. Table  \ref{tab:cloud details} shows the data availability. 
    
\end{enumerate}

\section{Results and Analysis}
Our goal is to compare the luminosities of \hcn\ and \hcop\ \jj10\ transitions to other conventional tracers of dense gas, in particular extinction thresholds and millimeter-wave emission from dust. We estimate the extinction from the \coo\ data and relations between gas column density and extinction. For the emission from dust, we use the mask file of the BGPS data, an indicator of the existence of gas of relatively high volume density \citep{2011ApJ...741..110D}. We also examine the ability of HCN, \hcop, and \coo\ to trace column density, determined from \textit{Herschel}  maps. 

 \subsection{Luminosity calculation based on Extinction criterion} \label{subsec:Av}
 
 This section describes the method of calculating luminosity of the line tracers coming from the dense regions and also from outside the dense region, depending on \coo\ column density based extinction(\av) map. It consists of two main steps: \textbf{(i)} making molecular hydrogen column density (\nh) maps of each target, convertible into maps of \av; and \textbf{(ii)} measuring the luminosity of regions satisfying the \av\ criterion.

 \subsubsection{Identifying the pixels with $\av \geq 8$ mag}
 
 To derive the molecular hydrogen column density maps of each target,  we use their \cotw\ and \coo\ data. We first identify the velocity range by examining the  spectrum  of \coo\ emission averaged over the whole mapped region of each target. The method of making column density maps of \coo\ is discussed in detail in \S Appendix \ref{app:H2}. We follow equation C1-C5 from \S Appendix \ref{app:H2} to produce the total column density map of \coo\ (N(\coo)) using \coo\ and \cotw\ data.   
 Next, we convolve and regrid the \coo\ column density map to match the TRAO map and convert it to a molecular hydrogen column density map using the following equation.
\begin{equation}
    \rm{N(H_{2})}=\rm{N(\coo)} \left[\frac{\rm{H_{2}}}{\rm{CO}} \right] \left[\frac{\rm{CO}}{\coo} \right]
\end{equation}
We assume a fractional abundance of 6000 for CO \citep{2017AAS...23021502L} and we apply the $[^{12}$C]/[$^{13}$C] isotopic ratio recently derived by \citet{2020A&A...640A.125J}. This ratio is linearly dependent on the \gal\ distance (\rg) of the target. 

\begin{equation}
    \frac{[^{12} \mbox{C}]}{[^{13}\mbox{C}]} = 5.87\times \frac{\rg}{\mathrm{kpc}} +13.25
\end{equation}
 Finally, we use the relation $\av= 1.0 \times 10^{-21}\nh $ to convert the molecular hydrogen column density to extinction \citep{2017AAS...23021502L}. This procedure allows us to define the pixels satisfying the criterion: $\av \geq 8$ mag.

 \subsubsection{The Emission from the Total region and Inside Region with $\av \geq 8$ mag}
 
 Next we calculate the average spectrum for two regions: 
 (1) including all pixels in the dense region and 
 (2) all pixels outside the dense region based on the \av\ criterion ($1.5 \le \av < 8 $ mag). The lower limit on \av\ is based on 5 times the RMS noise threshold in the \coo\ maps. Because the RMS noise levels are not the same for all the clouds  (see Table \ref{tab:cloud details}), we use  $\av = 1.5$ mag as the lower limit to make the analysis uniform.
 
\begin{figure*}
    \epsscale{1.2}
    \plotone{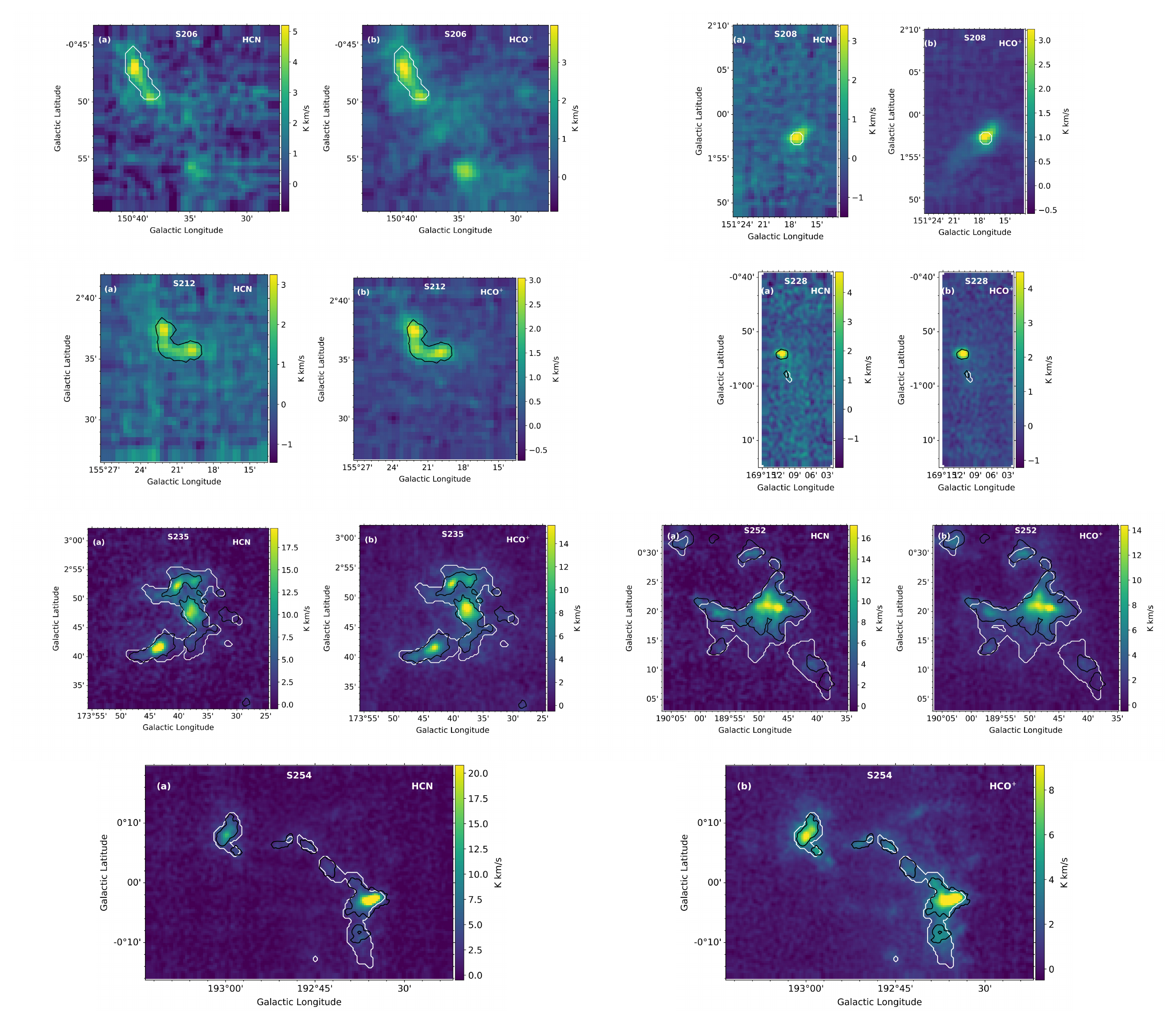}
    \caption{This figure shows the integrated intensity map of (a) \hcn\ and (b) \hcop\ for each target (name of each targets are mentioned in each image). The white contours indicate the $\av \ge 8$ mag regions and black contours indicate the BGPS mask regions. There are no black contours in \clouda\ and \cloudb\ as they lack BGPS data. There is no white contour in \cloudc, as there is no region above $\av=8$ mag.}
    \label{fig:panel}
\end{figure*}

\begin{figure*}
\gridline{\fig{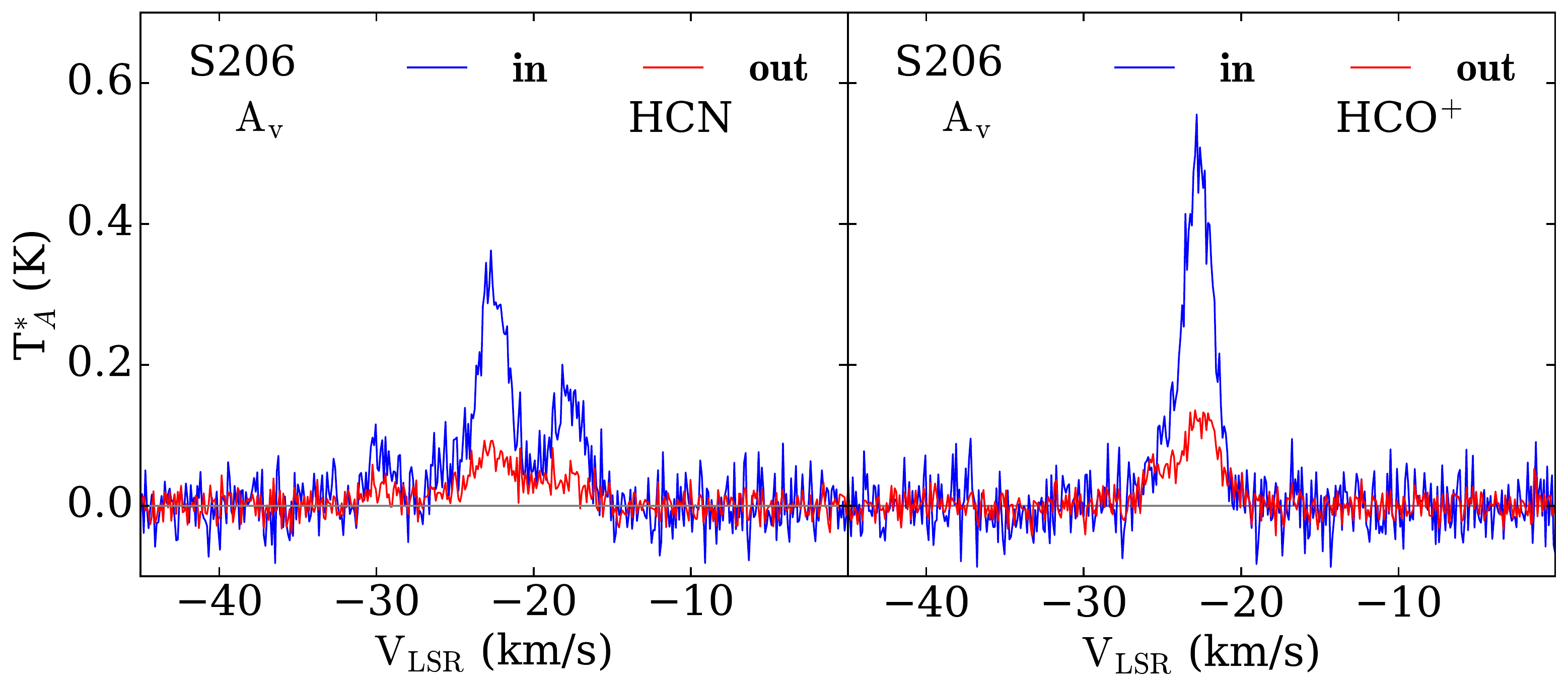}{0.38\textwidth}{(a)}
          \fig{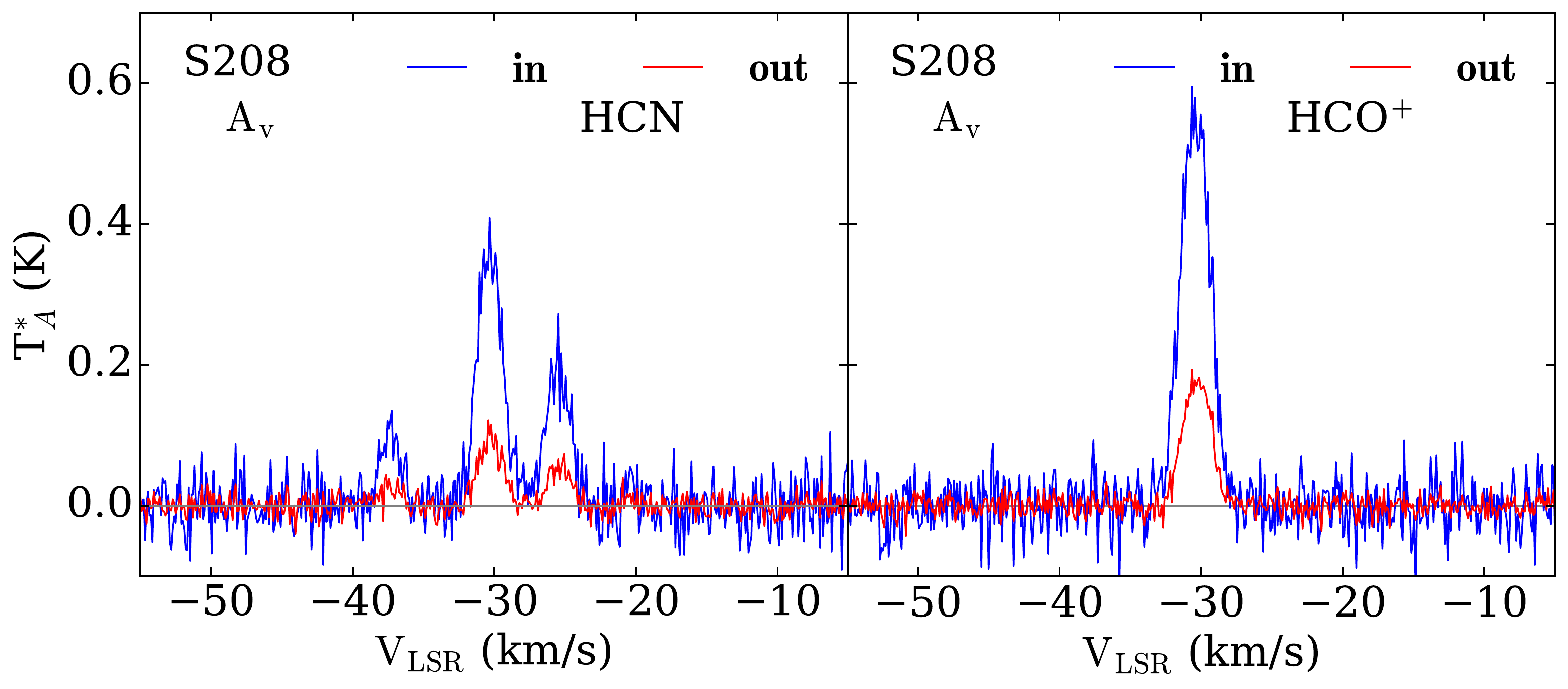}{0.38\textwidth}{(b)}
          }
\gridline{\fig{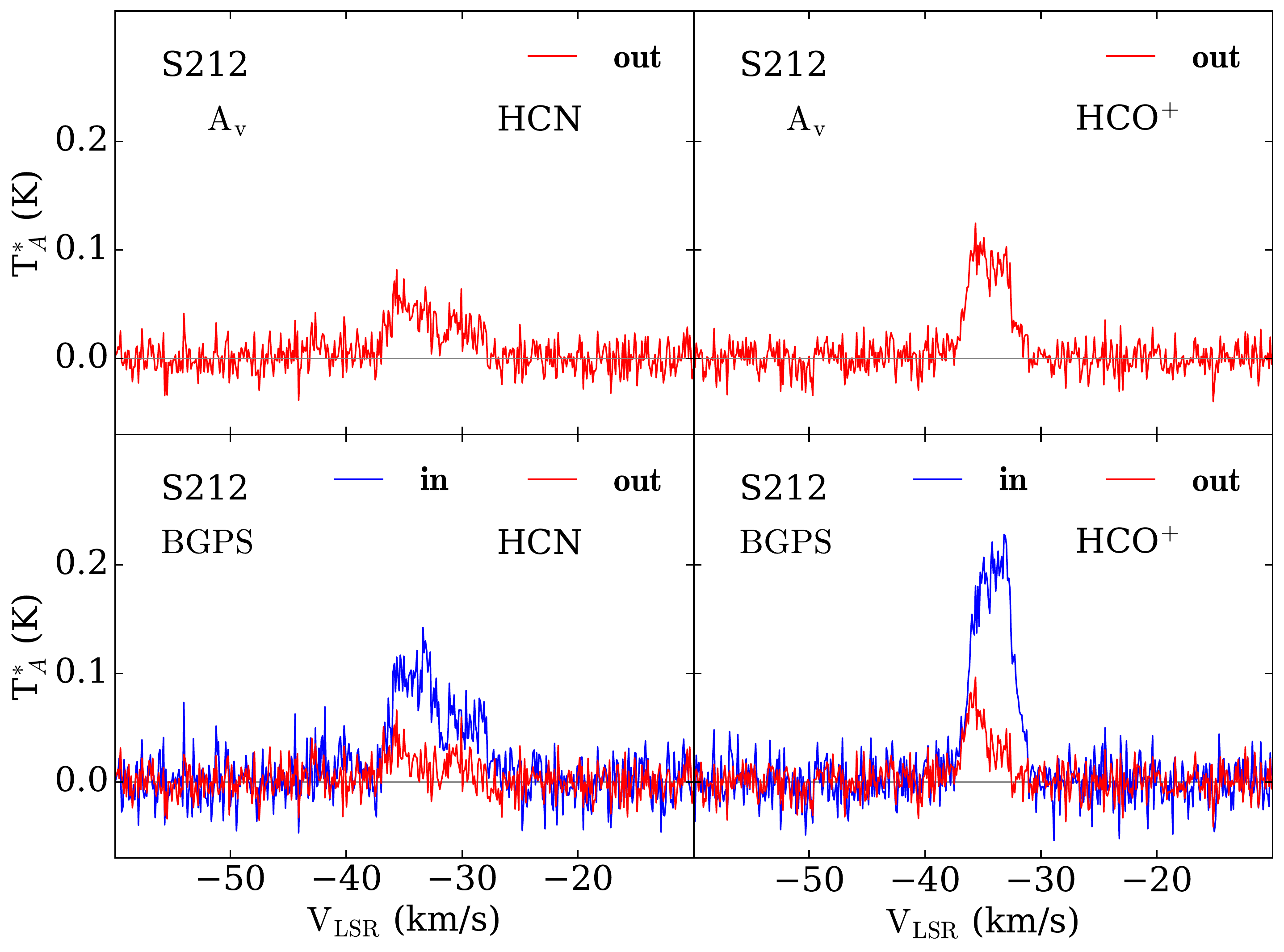}{0.38\textwidth}{(c)}
          \fig{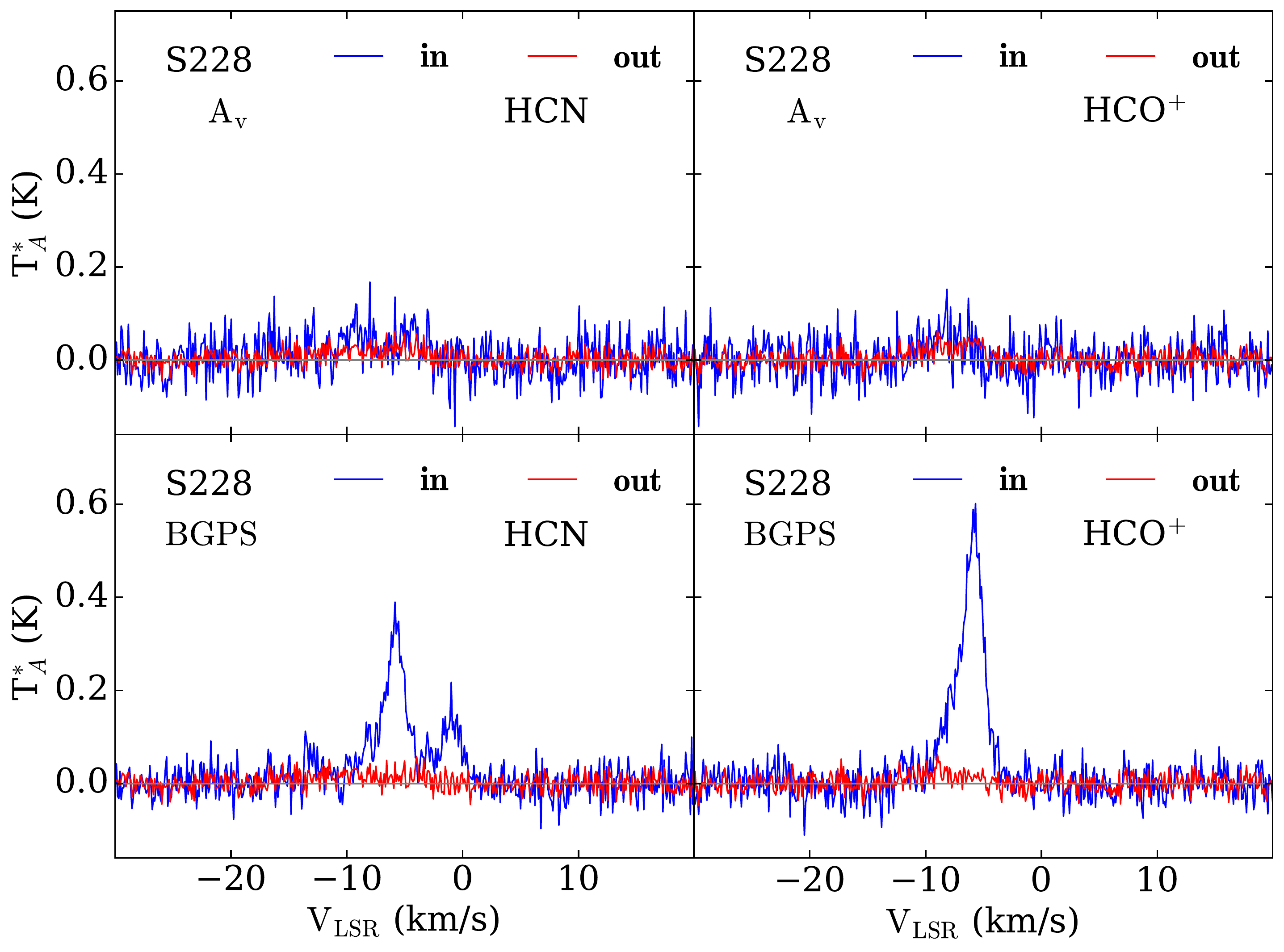}{0.38\textwidth}{(d)}
          }
\gridline{\fig{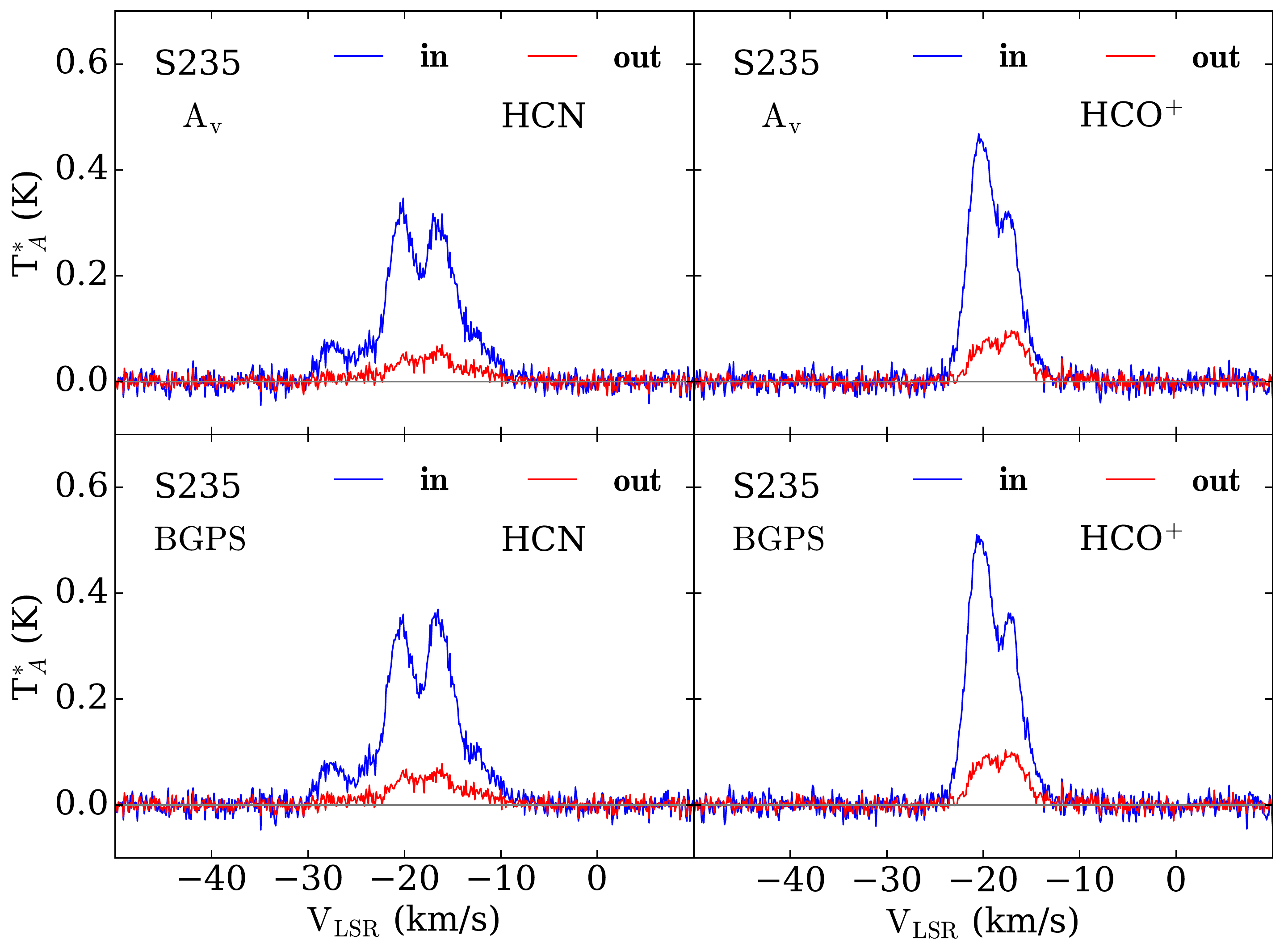}{0.38\textwidth}{(e)}
          \fig{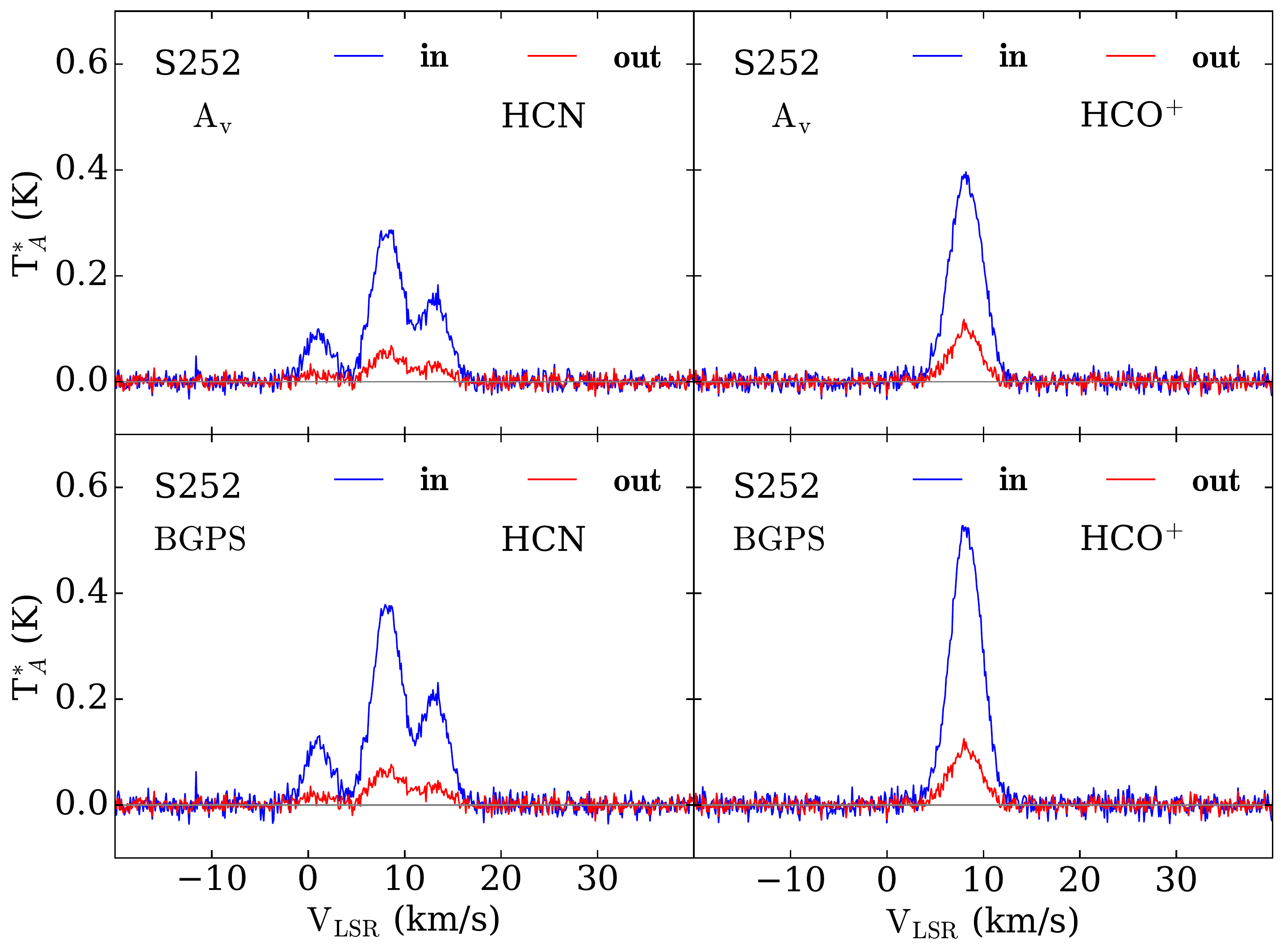}{0.38\textwidth}{(f)}
          }
\gridline{\fig{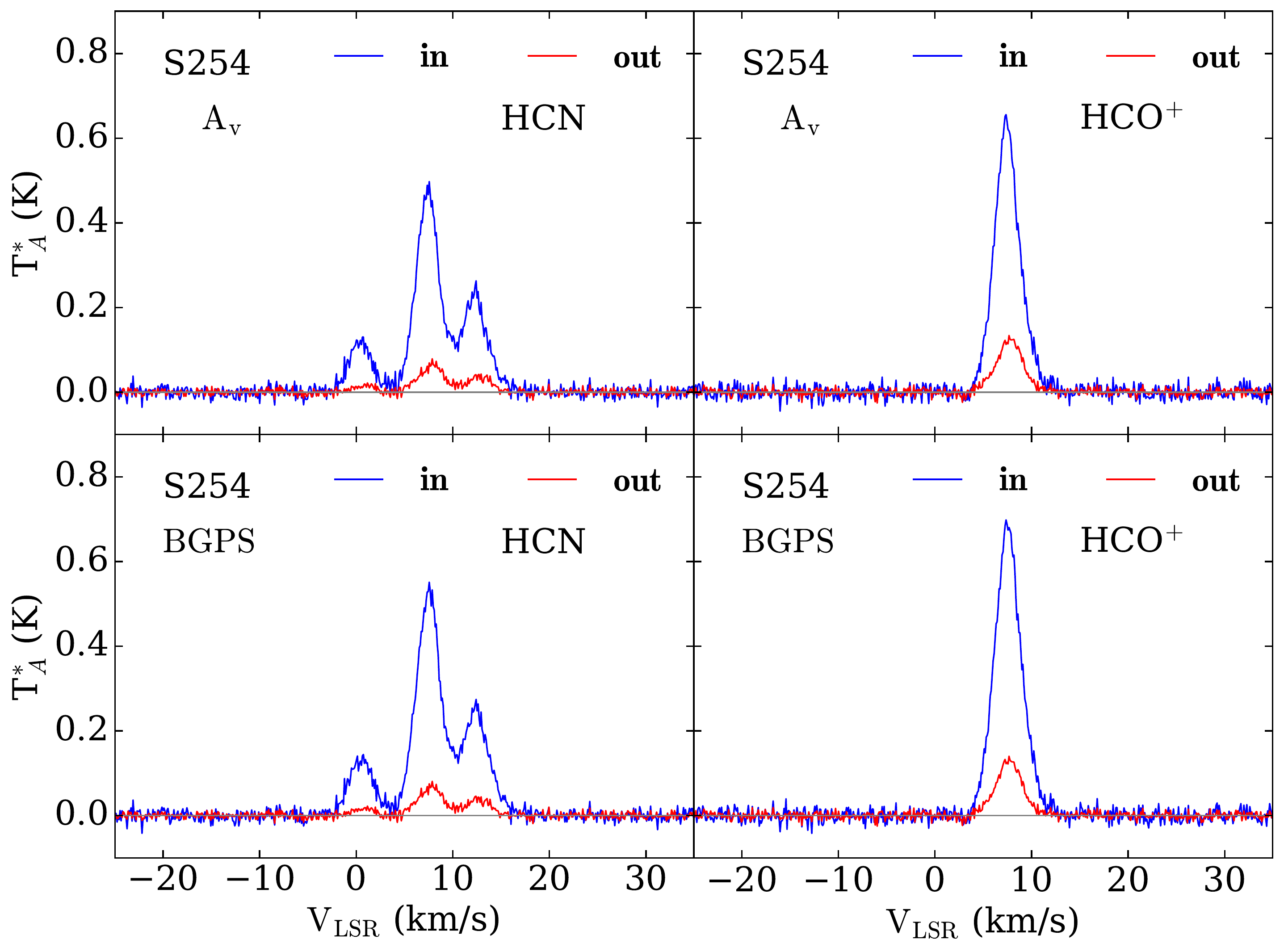}{0.38\textwidth}{(g)}}          
\caption{This is the average spectrum plot for the cloud (a)\clouda, (b)\cloudb, (c)\cloudc, (d)\cloudd, (e)\cloude, (f)\cloudf\ and (g)\cloudg\ based on \av\ criterion and BGPS mask map. The \textit{left} panel in all images are for \hcn\ and \textit{right} panels are for \hcop. 
The blue line indicates the line emission coming from $\av \ge8$ mag regions and red line indicates the emission from $1.5 \le \av < 8$ mag regions in \textit{top} panel of each sub-figure. In the \textit{bottom} panels of each sub-figures (except (a) and (b)), the blue line indicates emission coming from the BGPS masked region and red line is for emission outside the masked region.  
\label{fig:spectrum}}
\end{figure*}

 We first identify the dense regions with  $\av \ge 8\ \rm{mag}$ in the extinction map (denoted by \textbf{in}). Figure \ref{fig:panel} shows the integrated intensity map of \hcn\ and \hcop\ for all the targets; white contours on each image in Figure \ref{fig:panel} indicate the $\av \ge 8$ mag regions. The spectra averaged over pixels that satisfy the criterion $\av \ge 8\ \rm{mag}$ (blue line) and the spectra for pixels that do not (denoted \textbf{out} and shown in red line) are shown in Figure \ref{fig:spectrum} (a, b and upper panels of c, d, e, f, g). Figure (\ref{fig:spectrum}c) (\cloudc) does not show any blue line in the upper panel because there is no pixel satisfying the $\av \ge 8\ \rm{mag}$ criterion. The average spectral lines from the \textbf{in} region are stronger than those from the \textbf{out} region whenever both spectra are available. 
 One odd feature appears in Figure \ref{fig:panel} (\cloudd): the \hcn\ and \hcop\ agree well with the BGPS, but there is no peak in \av\ in that region. Figure (\ref{fig:spectrum}d) does not show strong emission for \hcn\ and \hcop\ at the position of $\av\ge8$ mag region (upper panel).
 This oddity and other information regarding individual targets are described in \S Appendix \ref{app:target}.

 For the calculation of luminosity, we use the main-beam corrected temperature of species $X$ ($T_{X}= T_{A}^{*}/\eta_{mb}$, where $\eta_{mb}=0.46$). 
 The velocity ranges for \coo\ and \hcop\ are similar and for \hcn\ we expand the velocity range to account for the hyperfine structure (around 3-6 \kms\ on each side depending on the target requirements). The  velocity ranges are explained in \S Appendix \ref{app:reduction}.

 We measure the luminosity of species $X$ arising from the $\av \ge 8\ \rm{mag}$ region using the following equation:
\begin{equation}
    L_{\rm X, in}=  D^2 \int_{v_{\rm sp, l}}^{v_{\rm sp, u}} dv \int d\Omega_{in} T_X(l,b,v),
\end{equation}
where $D$ is the distance of the target from the Sun in pc
and the integrals are taken over the line velocity between $v_{\rm sp, l}$ and $v_{\rm sp, u}$, the lower and upper limits of the line spectral window, and the solid angle that satisfies the extinction criterion.
In practice, we use the average spectrum in Figure \ref{fig:spectrum} to define the integrated intensity
\begin{equation}
    I_{\rm X} = \int_{v_{\rm sp, l}}^{v_{\rm sp, u}} T_X dv
\end{equation}
and compute
the luminosity from the following equation:
\begin{equation}
L_{\rm X, in}=  D^2  {I_{\rm X}} N_{\rm in} \Omega_{\rm pix},
\end{equation}
where $N_{\rm in}$ is the number of pixels satisfying the \textbf{in} condition and $\Omega_{\rm pix}$ is the solid angle of a pixel.

The uncertainty in luminosity is
\begin{equation}
    \sigma_{L} = D^{2} \delta v \sqrt{N_{ch}} \  \sigma_{T_X} \ N_{\rm in} \Omega_{\rm pix},
\end{equation}
where $N_{ch}$ is the number of channels in the line region, $\delta v$ is the channel width, and $\sigma_{T_X}$ is the RMS noise in the baseline of the average spectrum.

To calculate the total luminosity ($L_{\rm X,tot}$), we follow the same method but include all pixels with  $\av>1.5 \ \rm mag$ for $L_{\rm X,tot}$.

 We follow the same procedure for HCN, \hcop\ and \coo; for \coo, we made integrated intensity maps for the specified velocity range using the FITS files from FCRAO and these maps are convolved, resampled, and aligned with the TRAO maps.



 We tabulated the values for the fraction of pixels inside the \av\ criterion ($N_{\rm in}/N_{\rm tot}$), the log of total line luminosity (Log \ltot),  the log of the line luminosity coming from the $\av \ge 8 \ \rm mag$ region (Log \lin) and the fractional luminosity (\fl$=L_{in}/L_{tot}$) in Table \ref{tab:av8} for \hcn, \hcop and \coo. Also we have given the statistical parameters (mean, standard deviation and median) for the relevant columns. \cloudc\ has no pixels with $\av \ge8 \ \rm mag$.

\begin{figure}
    \epsscale{1.25}
    \plotone{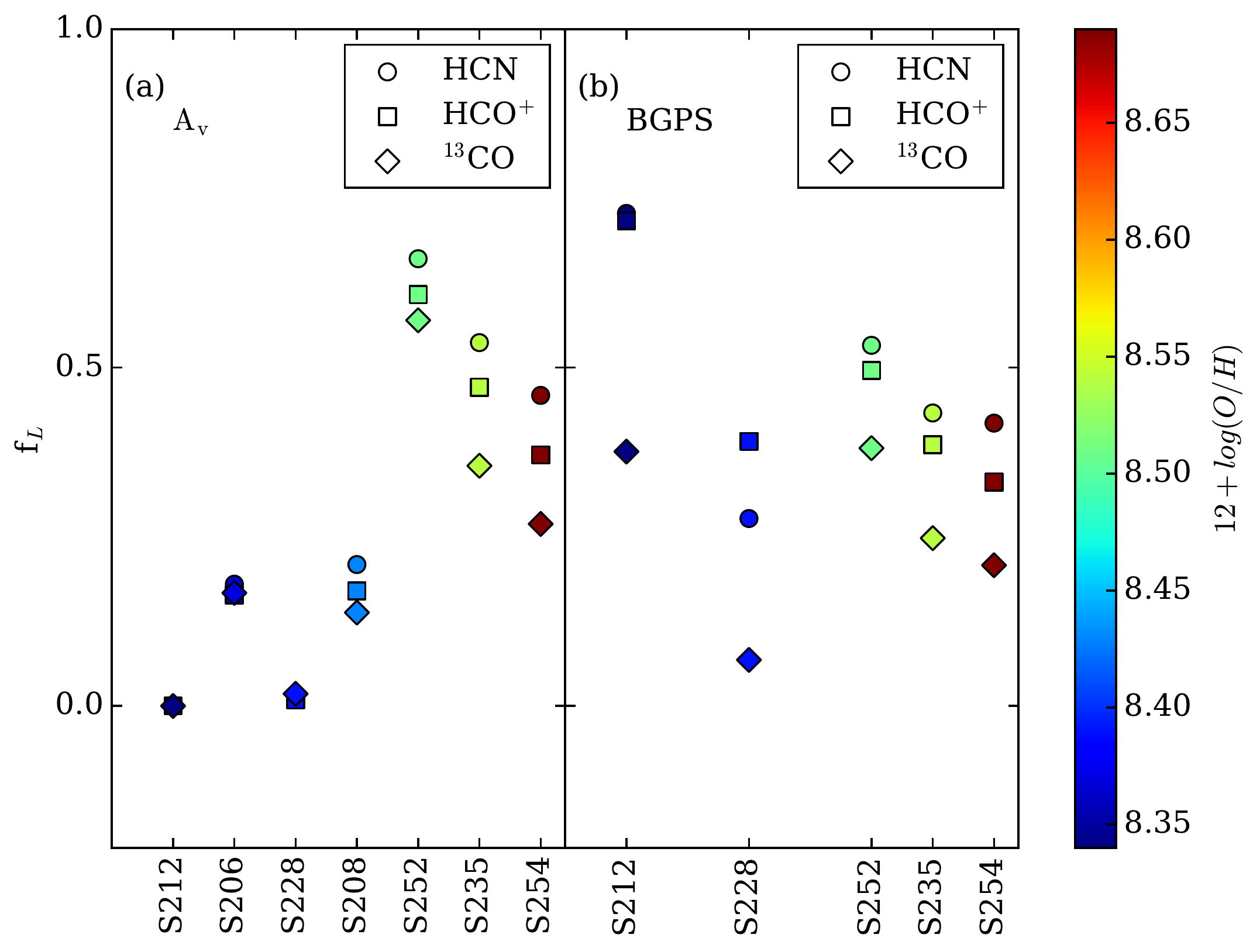}
    \caption{The fraction of luminosity (\fl) within \textbf{in} region is plotted for each tracer versus clouds. The \fl\ in the \textit{left} panel is obtained from $\av \ge8$ mag region and in the \textit{right} panel, \fl\ is from BGPS mask region. The targets are organized with increasing metallicity. Different symbols are used to separate the different tracers. Different color indicates the metallicity of each source.}
    \label{fig:fL_vs_cloud}
\end{figure}

 Figure \ref{fig:fL_vs_cloud}a shows \fl\ within the \textbf{in} regions defined by $\av \ge8$ mag (left) for each cloud sorted by increasing metallicity.
 The values of \fl\ vary between 0.014 and 0.661 for \hcn, between 0.009 and 0.608 for \hcop, and between 0.018 and 0.570 for \coo. The values of \fl\ for \hcn\ are higher than \hcop\ and \coo. 
 The mean value of the line luminosities coming from the \textbf{in} region is higher by 0.12 for \hcn\ compared to \hcop. For \coo, \lin\ is higher but the values of fractional luminosity (\fl) are lower than those of \hcn\ and \hcop. In terms of the logarithm of the total line luminosity, the mean is 2.78 for \coo, 1.98 for \hcop\ and 1.99 for \hcn. Comparing the two dense gas tracers, 
 \hcop\ and \hcn\ provide similar total luminosity,
 but \hcn\ shows better correlation with the extinction criterion in terms of \fl\ followed by \hcop\ and \coo\ (see Figure \ref{fig:fL_vs_cloud}a).   

\begin{figure}
    \epsscale{1.2}
    \plotone{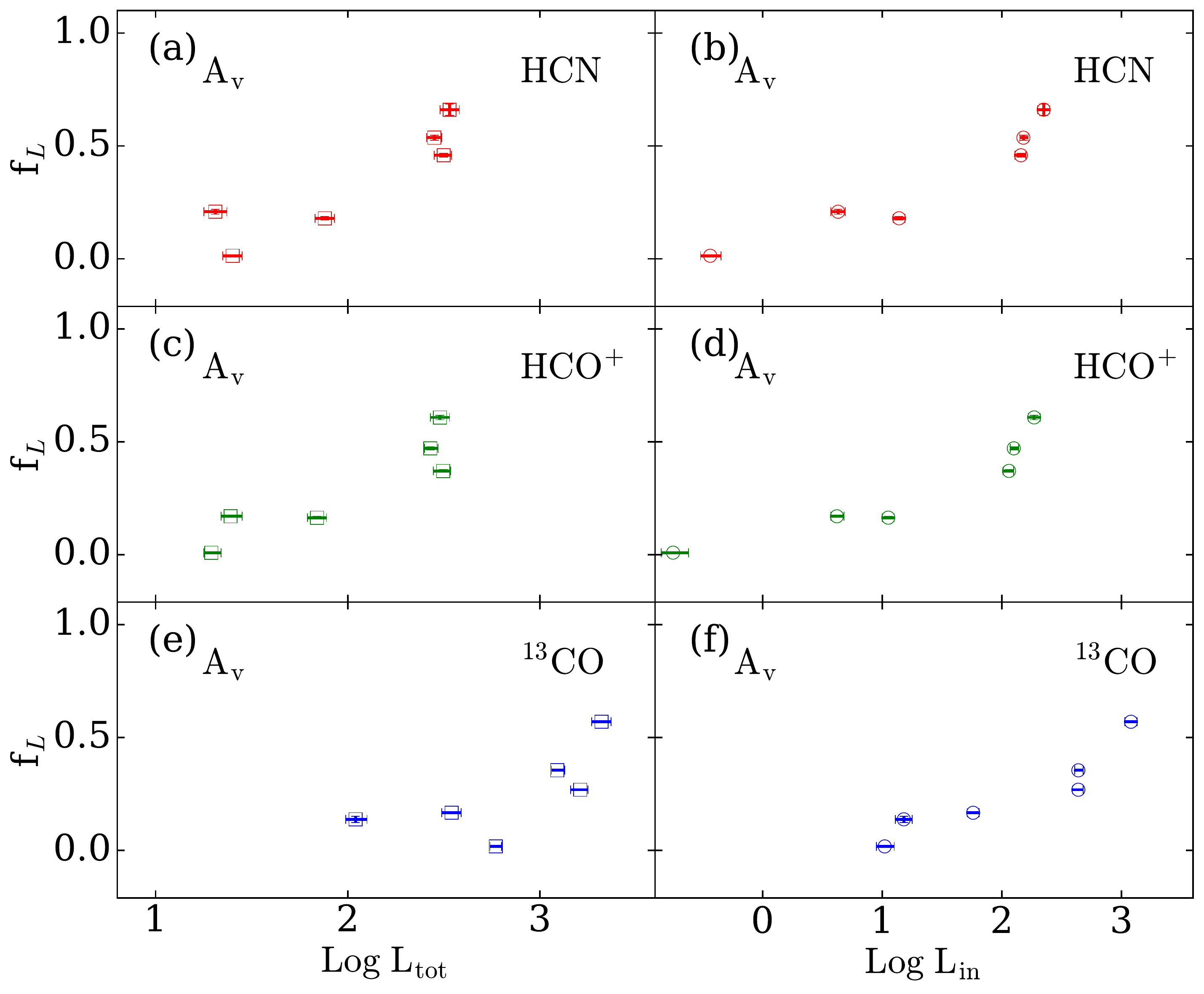}
    \caption{The fraction of line luminosity arising above $\av \ge 8$ mag is plotted versus line luminosity. 
    HCN is plotted on the \textit{top}, \hcop on the \textit{middle}, and \coo\ on the \textit{bottom}.
    The total line luminosities ($\mathrm{Log L_{tot}}$) are on the \textit{left} and luminosity from \textbf{in} region ($\mathrm{Log L_{in}}$) on \textit{right}}
    \label{fig:fL_vs_Luminosity}
\end{figure}

 The correlation between fractional line luminosity (\fl) arising from the $\av \ge 8$ mag region with $\mbox{Log}\ \ltot$ and $\mbox{Log}\ \lin$  
(Figure \ref{fig:fL_vs_Luminosity})  shows that the clouds with more material (i.e., higher luminosity) tend to have  higher values of \fl.

\subsection{Luminosity calculation from BGPS} \label{subsec:bgps}
 In this section, we follow the same basic method but use the dust thermal continuum emission to trace the `dense' gas. We use the data from BGPS to compare the line tracers in those regions. Millimeter continuum dust emission remains an optically thin tracer of the gas column density even in dense molecular clouds and efficiently identifies the molecular clumps that are possible sites of star formation. Because of the limits on sensitivity to large angular scale emission, the BGPS emission traces volume density, 
 but the characteristic density probed decreases with distance \citep{2011ApJ...741..110D}. For distances in this sample of 2-6 kpc, the mean density traced is 5\ee3 to 1\ee4 \cmv\ (Fig. 12 of \citealt{2011ApJ...741..110D}) with properties of clumps rather than cores or clouds. We use the mask maps of BGPS, downloaded from Bolocat V2.1 to define the \textbf{in} region. As before, we then  computed the line luminosities inside and outside the mask regions.

 First, we changed the BGPS mask maps to the same resolution and pixel size as the TRAO maps.  Next we make the average spectra of the \textbf{in} and \textbf{out} regions and use equation 5 to calculate the luminosity for \hcn, \hcop\ and \coo. The total luminosities (\ltot) are the same as those obtained using the extinction condition $\av \ge 8 \ \rm mag$ in Section \ref{subsec:Av}.  

 Among all the targets, 5 of the regions are covered by the BGPS survey. The black contours on the integrated intensity maps of \hcn\ and \hcop\ in Figure \ref{fig:panel} show the position of the BGPS \textbf{in} region.
 The bottom panels of Figure \ref{fig:spectrum}(c, d, e, f, g) show the average spectra of the luminosities coming from the BGPS \textbf{in} and \textbf{out} regions. All 5 targets show clear detection of \hcn\ and \hcop\ from the BGPS mask region, but significant emission is coming from outside the BGPS mask regions. 
 
 We present the result of those five targets in Table \ref{tab:bgps} along with the mean, standard deviation and median for relevant columns. We have listed the values for the fraction of pixels inside the BGPS mask region ($N_{\rm{in}}/N_{\rm{tot}}$), the log of the total line luminosity (Log $L_{\rm tot}$), the log of line luminosity arising from the BGPS \textbf{in} region (Log \lin) and the fractional luminosity (\fl$=L_{in}/L_{tot}$) in Table \ref{tab:bgps} for \hcn, \hcop and \coo.

 The mean \fl\ values, listed in Table \ref{tab:bgps}, are higher for \hcn, followed by \hcop\ and \coo\ (see Figure \ref{fig:fL_vs_cloud}b). 
 Dense line tracers (\hcn, \hcop) show better agreement with BGPS emission than they did with the extinction criterion ($\av \ge 8  \mathrm{mag}$) for these 5 clouds having BGPS data.  

\begin{figure}
    \epsscale{1.2}
    \plotone{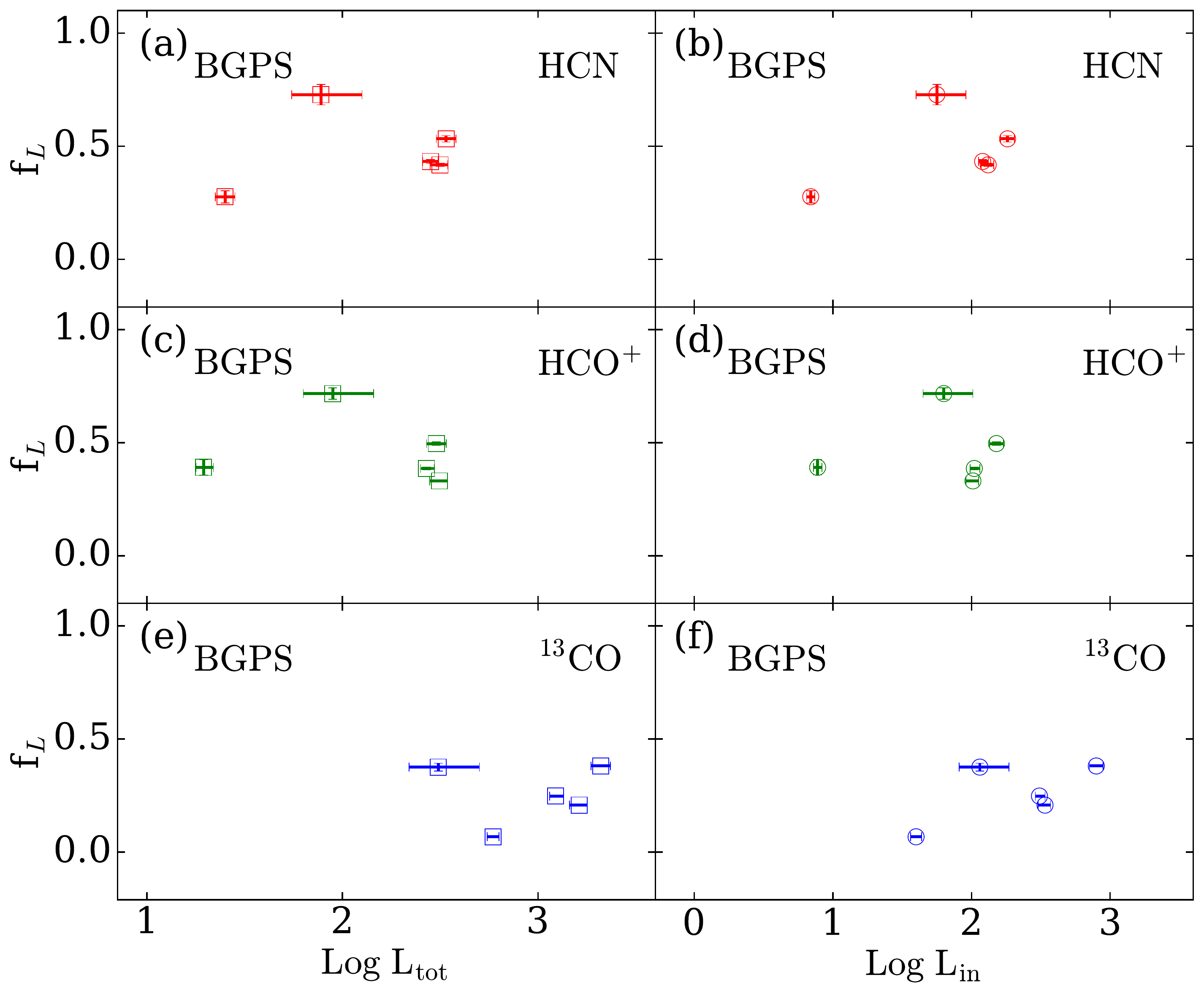}
    \caption{The fraction of line luminosity arising from the BGPS mask region is plotted versus line luminosity. 
    HCN is plotted on the \textit{top}, \hcop on the \textit{middle}, and \coo\ on the \textit{bottom}.
    The total line luminosities ($\mathrm{Log L_{tot}}$) are on the \textit{left} and luminosity from \textbf{in} region ($\mathrm{Log L_{in}}$) on \textit{right}.}
    \label{fig:fL_vs_Luminosity-bgps}
\end{figure}

 The standard deviation of the \fl\ values is 0.149 and 0.137 for \hcn\ and \hcop\ respectively, considerably lower than the values for the extinction criterion. Figure \ref{fig:fL_vs_Luminosity-bgps} shows the correlation between fractional line luminosity (\fl) arising from inside the BGPS mask region with $\mbox{Log} \lin$ and total luminosity $\mbox{Log}\ltot$. 
 The \fl\ values show no significant dependence on $\mbox{Log}\ \ltot$ and $\mbox{Log}\ \lin$ for any of the 3 tracers.

\subsection{Conversion of Line  Luminosity to Mass of Dense Gas} \label{subsec:Mdense}
In this section, we measure the mass conversion factor for the outer Galaxy targets. The conversion factor is defined as the ratio between the mass of gas and line luminosities ($M=\alpha_{Q} L_{Q}$, where the unit of $\alpha_{Q}$ is $\mathrm{\msun (K \ km \ s^{-1}\ pc^{2})^{-1}}$. This conversion factor is used in extragalactic studies to estimate the mass of dense gas.

The dense gas mass of each source having BGPS data is estimated from its integrated flux density, $S_{1.1}$ (available in BGPS Source catalog table\footnote{\url{https://irsa.ipac.caltech.edu/data/BOLOCAM_GPS/tables/bgps_v2.1.tbl}}, \citealt{2010ApJS..188..123R}) combined with distance ($D$) and dust temperature ($T_{dust}$). The 1.1 mm dust emission is assumed to be optically thin and  at a single temperature, and a dust opacity is assumed.     

\begin{equation}
    \mdense = \frac{S_{1.1}D^{2}(\rho_{g}/\rho_{d})}{B_{\nu}(T_{dust})\kappa_{dust,1.1}}
\end{equation}
where $B_{\nu}$ is the Planck function evaluated at $\lambda=1.1$ mm, $\kappa_{dust,1.1}=1.14$ $\rm{cm^2 g^{-1}}$ is the dust opacity per gram of dust \citep{2006ApJ...638..293E} and gas-to-dust ratio ($\rho_{g}/\rho_{d}$) is assumed to be 100 \citep{1983QJRAS..24..267H}. 
We use the relation after scaling it \citep{2010ApJS..188..123R}

\begin{equation}
    \mdense = 13.1 M_{\odot}\left(\frac{D}{1 \mathrm{ kpc}}\right)^{2}\left(\frac{S_{1.1}}{1 \mathrm{Jy}}\right)
    \left[ \frac{e^{(13.0 \mathrm{K}/T_{dust})}-1 }{e^{(13.0/20)}-1} \right]
\end{equation}
For consistency with \citet{2010ApJS..188..123R}, we assume a single dust temperature ($T_{dust}=20$ K) for all sources to estimate the dense gas mass.  For the sources with Herschel data, the derived $T_{dust}$ is consistent with this choice. The values of dense mass (\mdense) are listed in Table \ref{tab:Mdense} in logarithmic scale. The uncertainties include distance uncertainties. 


The mass conversion factor \at\ is calculated as the mass of dense gas divided by the total luminosity of the clouds ($\av > 1.5$ mag) ($\at = \mdense/\ltot$) and for \ain\ the luminosity is restricted to the dense region based on BGPS mask ($\ain=\mdense/\lin$).  
In Table \ref{tab:Mdense}, we list the conversion factors along with total line luminosities (\ltot), line luminosities inside the BGPS mask (\lin) in logarithmic scale for \hcn\ and \hcop. We also estimated the statistical parameters (mean, standard deviation and median) to understand the correlation between \mdense\ and the conversion factors (\ain, \at).  
The mean value of \ain\ in logarithmic scale is $\langle \mathrm{Log \ \ain(\hcn)} \rangle = 1.28\pm 0.08$, translating to $\mdense= 19 ^{+3.8}_{-3.2} \lin(\hcn)$ and for \at,  $\langle \mathrm{Log \ \at(\hcn)} \rangle = 0.93\pm 0.16$, translating to $\mdense= 8.5^{+3.8}_{-2.6} \ltot(\hcn)$.  
Our \at\ value is close to the extragalactic conversion factor ($\alpha_{\rm{HCN(1-0)}} = 10$ \msun$\mathrm{(K\ km\ s^{-1}\ pc^{2})^{-1}}$, \citet{2004ApJS..152...63G}), where luminosity from the whole cloud is considered to obtain mass. 
The \ain\ value is similar to the value obtained by \citet{2010ApJS..188..313W} for dense clumps. 
The theoretical prediction of $\mathrm{\alpha(HCN)}$ by \citet{2018MNRAS.479.1702O} is $14\pm6$ $\mathrm{\msun(K\ km \ s^{-1}\ pc^{2})^{-1}}$ for $n \geq$1\ee4 \cmv\ which corresponds to the dense gas of the cloud. Including the uncertainty, this theoretical value is consistent with either our \ain(\hcn) or \at(\hcn).
Calculations for \hcop\ yields $\mathrm{\ain= 20.4^{+4.1}_{-3.4}}$  $\mathrm{\msun(K\ km \ s^{-1}\ pc^{2})^{-1}}$ and $\mathrm{\at= 9.1^{+3.2}_{-2.3}}$  \msun$\mathrm{(K\ km\ s^{-1}\ pc^{2})^{-1}}$, similar to the values for \hcn.

Variation in $T_{dust}$ by $\pm3$ K produce changes in \ain\ and \at\ within the variation among sources, whereas variation by $5$ K in dust temperature begin to produce changes outside that range. Higher $T_{dust}$ of course produce lower values of \ain, \at; lower $T_{dust}$ gives higher values.

We also estimate the mass of BGPS clumps from the Herschel based column density maps from \citet{2017MNRAS.471.2730M} for \cloudd, \cloudf\ and \cloudg. We use $\mathrm{M_{dense} = \mu_{H_{2}} m_{H} A_{pix} \sum N(H_{2})}$, where mean molecular weight $\mathrm{\mu_{H_{2}}}$ is assumed to be 2.8 \citep{2008A&A...487..993K}, $\mathrm{m_{H}}$ is mass of hydrogen, $\mathrm{A_{pix}}$ is area of a pixel in $\mathrm{cm^{2}}$ and $\mathrm{\sum N(H_{2})}$ is the integrated column density.
The mass values in logarithmic scale are 2.36, 4.09 and 3.72 respectively for 3 targets. The values are higher than BGPS mass measurement by average 0.34 in log scale, which is 2.2 in linear scale. The difference in the values is mainly caused by differing assumptions about the dust opacity. \citet{2017MNRAS.471.2730M} used $\mathrm{\kappa(\lambda)=0.1 (\frac{\lambda}{300})^{-2}}$ $\mathrm{cm^{2}g^{-1}}$, which implies a dust opacity per gram of dust at 1.1 mm of $\mathrm{\kappa_{1.1}}=0.74$ $\mathrm{cm^{2}g^{-1}}$, a value  1.5 times lower than the $\mathrm{\kappa_{dust,1.1}}$ value assumed in the BGPS mass measurement. 

BGPS resolves out the extended structure of the clouds, it gives a better measure of mean volume density of the dense regions than Herschel data. Our aim is to measure the L(\hcn) in the mean high volume density region, not the column density region, for which Herschel works best.

\begin{figure*}
\gridline{\fig{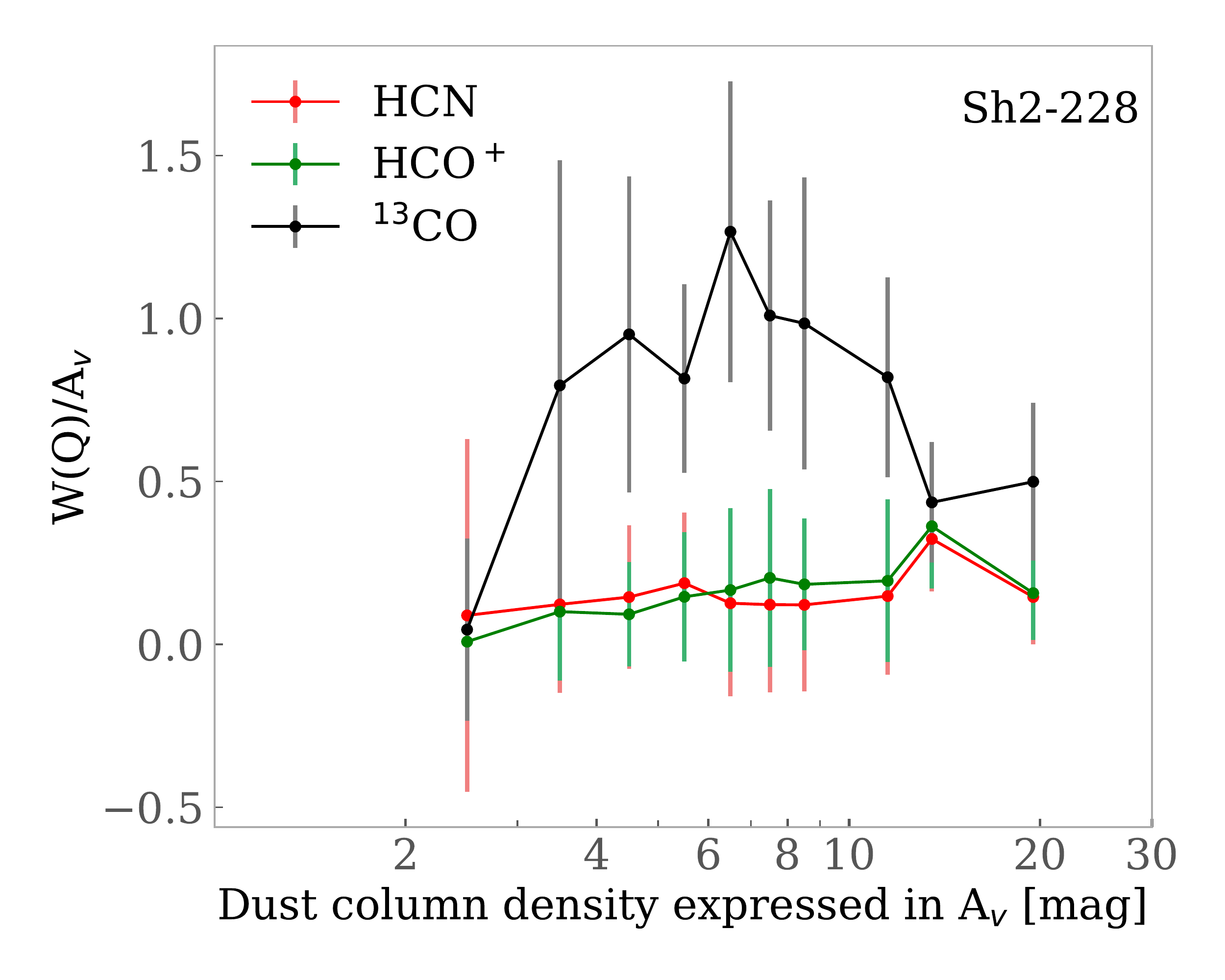}{0.35\textwidth}{(a)}
          \fig{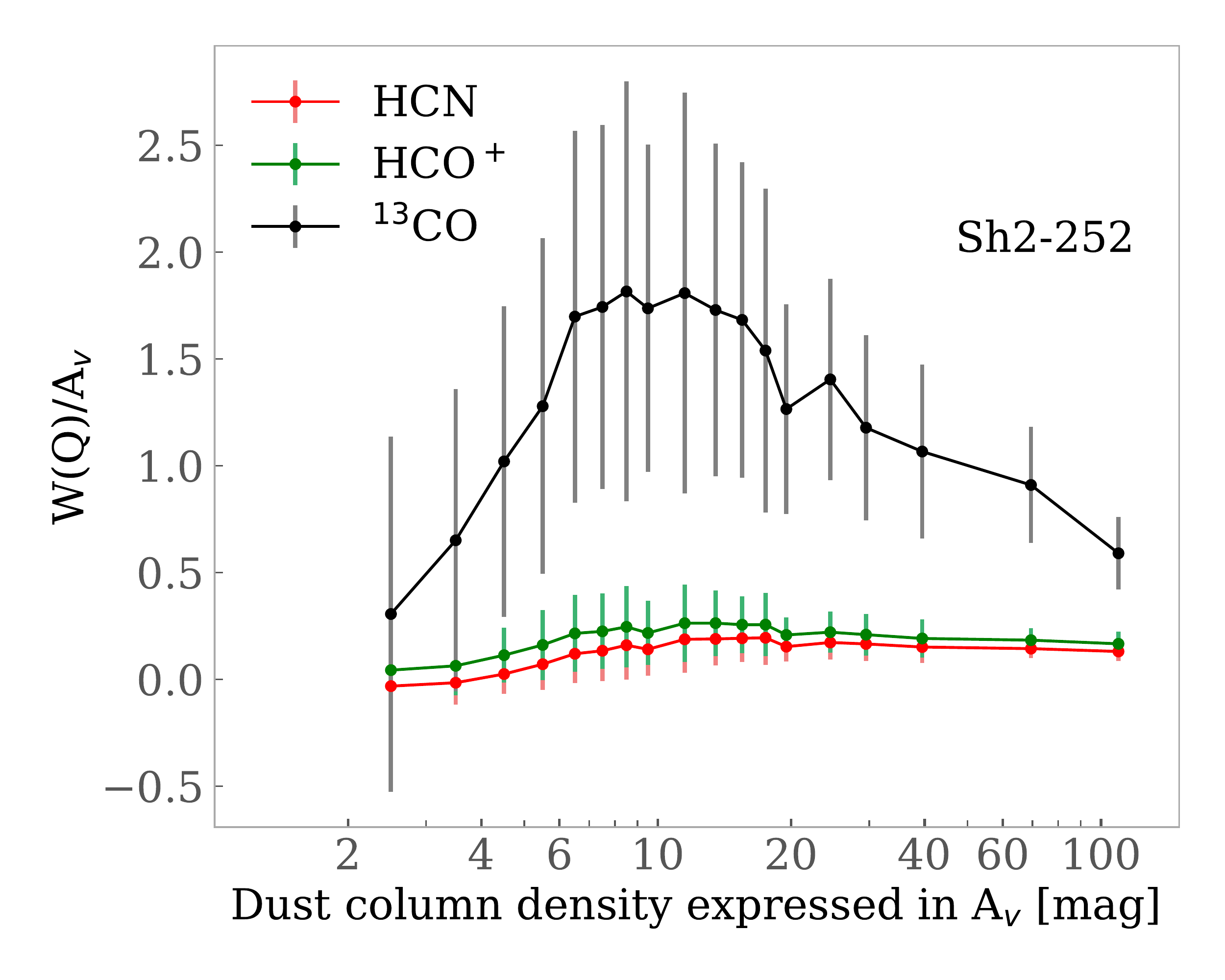}{0.35\textwidth}{(b)}
          \fig{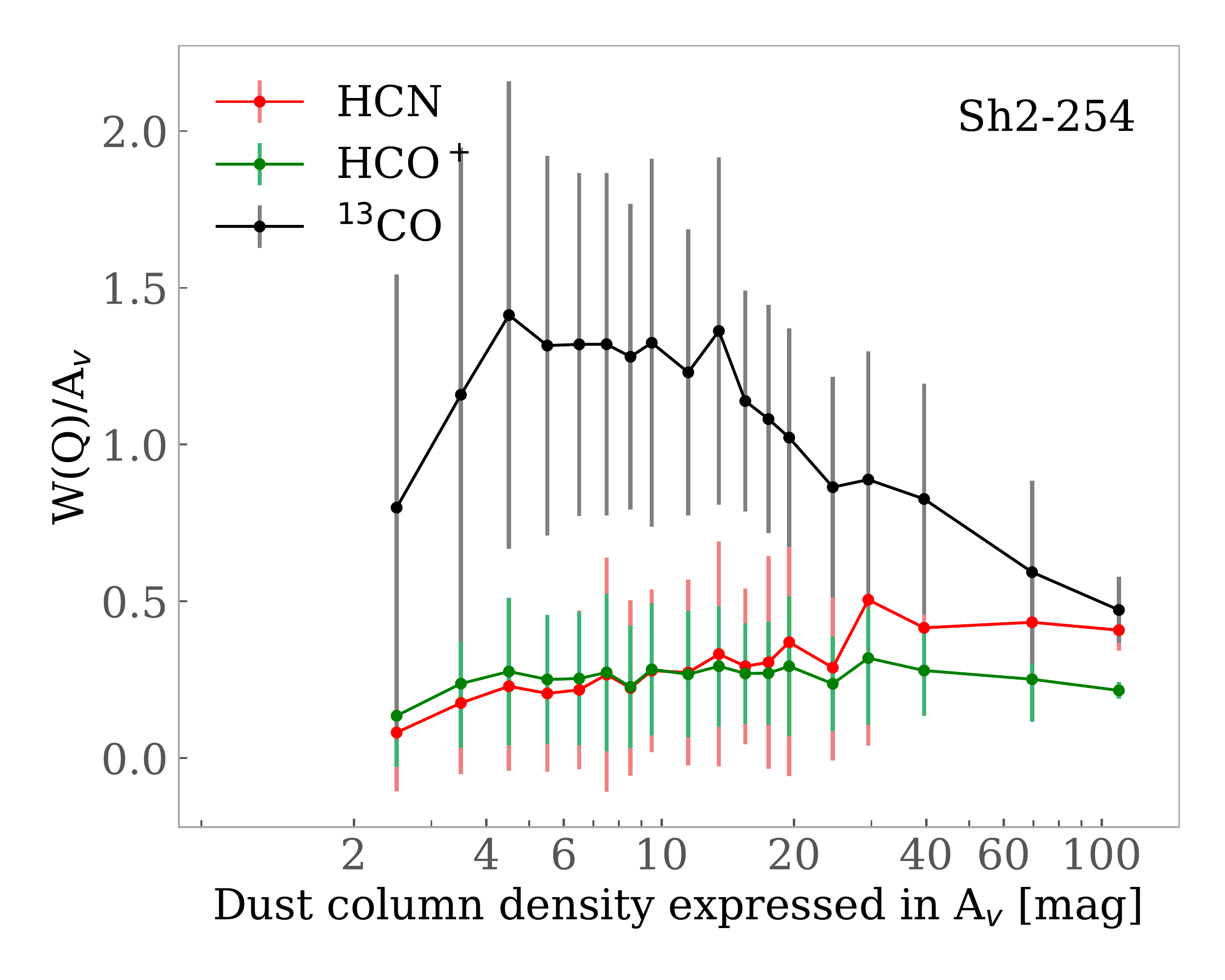}{0.35\textwidth}{(c)}
          }          
\caption{Line-to-mass ratio is plotted against \av\ based on Herschel dust column density for the region (a) \cloudd, (b) \cloudf\ and (c) \cloudg.
\label{fig:herschel}}
\end{figure*}

\subsection{Line Tracers versus Herschel Column density} \label{subsec:Herschel}
 We also explore how well  molecular lines trace the dust column density derived from Herschel data.  \citet{2017A&A...605L...5K} and \citet{2020MNRAS.497.1972B} studied the relationship between several molecular line emissions and dense gas in Orion A and W49 molecular clouds. 
 The line-to-mass ratio is defined by $h_{Q}=W(Q)/\av$, where $W(Q)=\int T_{mb}dv$ and $\av \propto \mathrm{N(H_{2})}$. 
 This ratio shows the relation between molecular line emission from transition $Q$ and the mass reservoir characterized by \av\ \citep{2017A&A...605L...5K}.
 The factor $h_{Q}$ is considered as a proxy for the line emissivity, or the efficiency of an emitting transition per \hh\ molecule \citep{2020MNRAS.497.1972B}.
 
 We plot line-to-mass ratio ($h_{Q}$) versus extinction (\av) based on Herschel column density for the 3 targets for which we have data (see Table \ref{tab:cloud details}). We have used the Herschel column density data from \citet{2017MNRAS.471.2730M} for \cloudd, \cloudf, \cloudg\  after regridding and convolving to TRAO maps resolution and pixel size. Figure \ref{fig:herschel} shows how $W(Q)/\av$ varies with increasing \av\ obtained from dust based column density. We exclude pixels with $\av < 2$ mag because of observational uncertainties.

\begin{figure}
    \epsscale{1.2}
    \plotone{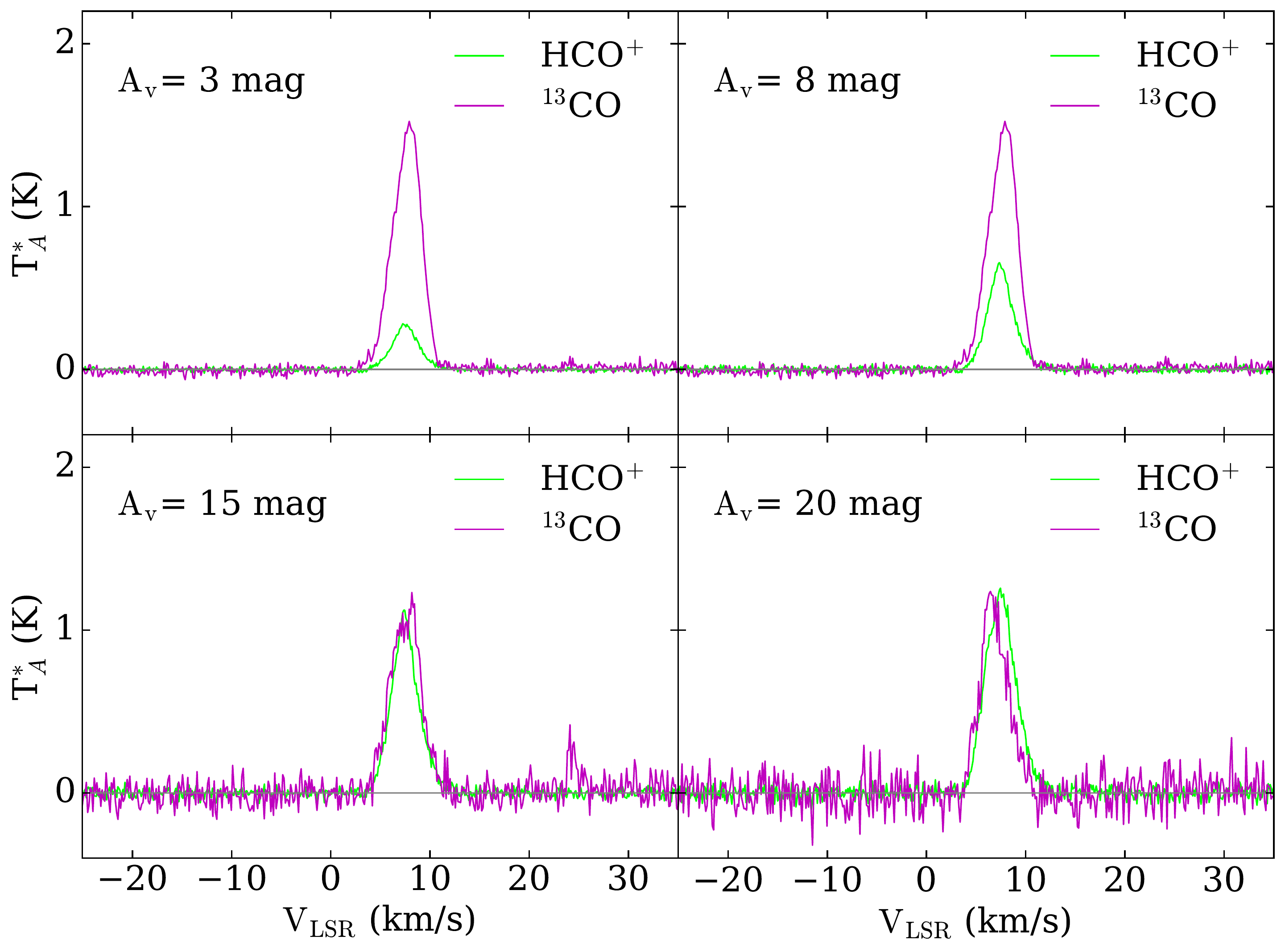}
    \caption{\hcop\ and \coo\ spectra are plotted for \cloudg\ with increasing \av\ value for \cloudg. \coo\ becomes optically thick in higher column densities.}
    \label{fig:optical_depth}
\end{figure}

 The line emissivity ($h_{Q}$) of \coo\ increases up to $\av \approx 6-7$ mag and then decreases gradually with increasing \av, whereas for \hcop\ and \hcn\ it does not  decrease. \coo\ traces the diffuse part well but fails in high column density regions, where \hcn\ and \hcop\ are more reliable.  Both \fl\ and $W(Q)/\av$ indicate that dense gas is traced best by \hcn\ and \hcop\ but less well by \coo. 
 Two possible reasons for \coo\ falling at high \av\ are the following: (i) it becomes optically thick at high column densities  or (ii) if dust temperature falls below 20-25 K, \coo\ depletes on dust grains \citep{1997ApJ...486..316B}. 
 Figure \ref{fig:optical_depth} explores the first reason by plotting the spectra of \coo\ and \hcop\ with increasing \av\ for \cloudg.  \hcop\ continues to grow stronger with increasing column density while \coo\ actually becomes somewhat weaker. Although no self-absorption features are prominent, the \coo\ line is likely optically thick. A similar effect has been seen in the other 2 targets as well. In principle, the optical depth is corrected for in the LTE model, but that model can fail if the excitation temperatures of the \cotw\ and \coo\ differ, as we discuss in \S Appendix \ref{app:H2} in more detail. Fortunately, the failure of \coo\ to trace column density mostly arise at $\av > 8$ mag, so that our method using \coo\ to identify the dense regions is still valid.

\section{Discussion} \label{sec:discussion}



\subsection{Comparison with Inner Galaxy targets\label{subsec:innercomp}}
A preliminary comparison between the outer and inner Galaxy targets can provide hints about the effects of Galactic environment. In their study of
 6 inner Galaxy clouds with \hcn\ and \hcop (\jj10), \citet{2020ApJ...894..103E} found a significant amount of luminosity from the diffuse part of the clouds. That is also true for the outer Galaxy clouds, but the fraction of luminosity arising inside the $\av \geq 8$ mag or the BGPS region is larger in the outer Galaxy clouds. Also, the behavior of individual tracers differ in both regimes. For inner Galaxy clouds,
 the average ratio $\langle \fl(\hcop)/\fl(\hcn) \rangle = 1.14\pm 0.18$ for $\av \ge 8 \ \mathrm{mag}$ criterion compared to $0.83\pm 0.10$ for the outer Galaxy clouds. 
For  inner Galaxy clouds \fl(\hcop) is greater than \fl(\hcn), but for the outer Galaxy the order is opposite.      

\begin{figure}
    \epsscale{1.2}
    \plotone{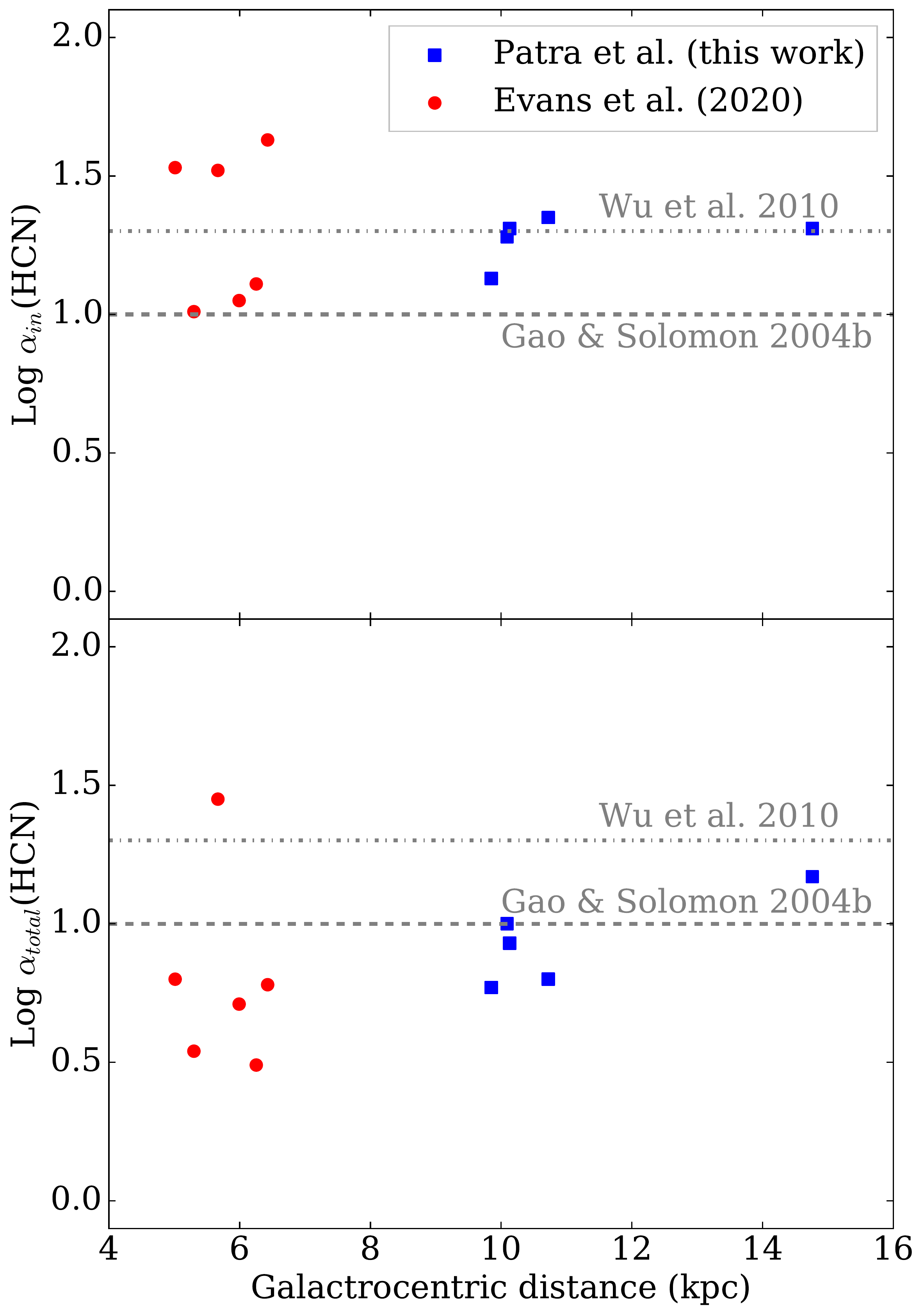}
    \caption{Variation of conversions factors $\mathrm{Log\ \ain(\hcn)}$ and $\mathrm{Log\ \at(\hcn)}$ for inner and outer Galaxy targets with \gal\ distance(\rg).}
    \label{fig:conversionfactor}
\end{figure}

\citet{2020ApJ...894..103E} also calculated the conversion factor between \mdense\ measured by millimeter-wave continuum data and luminosity coming from the dense region. For inner Galaxy clouds \ain(\hcn) is $20^{+16}_{-9}$ $\mathrm{\msun (K \ km \ s^{-1}\ pc^{2})^{-1}}$, whereas it is $19^{+3.8}_{-3.2}$ $\mathrm{\msun (K \ km \ s^{-1}\ pc^{2})^{-1}}$ for outer Galaxy clouds. Considering the whole cloud, the mass-luminosity conversion factor, \at(\hcn) was $6.2^{+6.6}_{-3.0}$ $\mathrm{\msun (K \ km \ s^{-1}\ pc^{2})^{-1}}$ for inner Galaxy clouds and \at(\hcn) is $8.5^{+3.8}_{-2.6}$ $\mathrm{\msun (K \ km \ s^{-1}\ pc^{2})^{-1}}$ for outer Galaxy clouds.
From the above values, we can see that \at\ and \ain\ for outer Galaxy clouds are consistent with inner Galaxy values. 
Figure \ref{fig:conversionfactor} shows the variation of conversion factors with \rg\ for both the set of targets. 
However, the mean density probed depends on heliocentric distance, and the inner Galaxy clouds are on average farther away. The high outlier in the lower panel of Figure \ref{fig:conversionfactor} is an inner Galaxy cloud that had very strong self-absorption in \hcn, lowering the line luminosity and raising \at(\hcn).  Bigger samples are needed to understand all the effects.

\subsection{Potential Impact of metallicity on fractional Luminosity?\label{subsec:metallicity}}
 
 \hcn\ and \hcop\ are linear molecules with large dipole moments and similar energy levels, but differing in the elemental compositions ($\mathrm{N}$ versus $\mathrm{O}$). This difference in chemical structure makes them sensitive to the chemistry of the interstellar medium. 
 Hence we can expect a connection between the abundance ratio of $\mathrm{N/O}$ and  $\mathrm{\hcn/\hcop}$ \citep{2017A&A...597A..44B}.
 The ratio of total luminosities, $L(\hcop)/L(\hcn)$, averaged in the logs, is 0.98 for outer Galaxy clouds, while this factor is 0.56 for inner Galaxy targets \citep{2020ApJ...894..103E}. The higher value in the outer Galaxy is consistent with the extragalactic studies on low metallicity systems (1.8-3, LMC, SMC \citep{1997A&A...317..548C, 2020A&A...643A..63G}; 1.1-2.8, IC 10 \citep{2016ApJ...829...94N, 2017A&A...597A..44B, 2018ApJ...862..120K}; 1.1-2.5, M33 \citep{2013A&A...549A..17B, 2017A&A...597A..44B}; 1.2, M31 \citep{2005A&A...429..153B}). 
 So, in the low metallicity environments \hcop\ is  as luminous as \hcn\ but less concentrated in dense regions compared to high metallicity regions.

 The \fl\ values based on \av\ analysis increase with the metallicity (Figure \ref{fig:fL_vs_cloud}), but  \fl\ values based on BGPS analysis do not.
 A possible reason for this difference is that we use $\mathrm{[\cotw]/[\coo]}$ ratio in the derivation of molecular hydrogen column density ($\mathrm{\nh}$) map (see \ref{subsec:Av}), which is a function of \rg\ and indirectly related to metallicity. For BGPS, the same opacity value is used regardless of metallicity, so the metallicity effect is not incorporated there. This study hints at interesting trends with Galactic environment and indicates the need for a larger sample to study it further.  

\section{Conclusions} \label{sec:conclusion}

In this paper, we study 7 outer Galaxy star-forming regions with \hcn\ and \hcop\ \jj10 transitions obtained from TRAO 14-m telescope and compare these dense line tracers with extinction, 1.1 mm thermal dust continuum emission (BGPS), and dust column density from Herschel.  The results are summarized below.

\begin{enumerate}
    \item The luminosity coming from the `dense' region based on extinction criteria and BGPS mask maps (indicated as \textbf{in}) is prominent for most of the clouds, but there are significant amounts of luminosity coming from outside the `dense' region (indicated as \textbf{out}) for most of the clouds. 
    The fraction of the total line luminosity arising from the dense  region, as indicated by both the BGPS and extinction maps, is higher in \hcn, followed by \hcop\ and \coo. The \hcn\ emission is generally less extended than the \hcop\ emission. 
    
    \item  In the outer Galaxy, \hcop\ is as luminous as \hcn,  but \hcop\ is less concentrated in the dense regions; these are the opposite of the trends in the inner Galaxy.

    \item Both \hcn\ and \hcop\ show better agreement with millimeter continuum emission than they do with the extinction criterion for the clouds having BGPS data. The dense line tracers in \cloudc\ and \cloudd\ do not agree with the extinction criterion.
    
    \item The fraction of line luminosity arising from the dense region based on extinction criterion increases with metallicity, but no such variation for the analysis based on millimeter continuum (Figure \ref{fig:fL_vs_cloud}). 
    For lower metallicity targets, the  fraction of the luminosity of the three tracers (\hcn, \hcop, and \coo) arising in high extinction regions are comparable, but all are low.  
    
    \item For 5 clouds, we estimate the mass conversion factor (\ain) between dense gas mass (\mdense) arising from BGPS mask region and the line luminosities of \hcn, \hcop, both within that region (\ain) and for the whole cloud (\at) in \S \ref{subsec:Mdense}. 
    The \at\ value is consistent with the literature value used in extragalactic studies ($10\ \msun \ (\rm K\ km\ s^{-1}\ pc^{2})^{-1}$, \citet{2004ApJS..152...63G}), while the \ain\ value is consistent with those found in studies of dense Galactic clumps \citep{2010ApJS..188..313W}. Also, the theoretical value obtained by \citet{2018MNRAS.479.1702O} for \hcn\ is consistent with either our \ain(\hcn) or \at(\hcn).
      
    \item We measure the line-to-mass ratio ($W(Q)/\av$) for the 3 targets (\cloudd, \cloudf\ and \cloudg) with column densities from Herschel (\S \ref{subsec:Herschel}). The \coo\ traces column density well over the range $\av = 4-10$ mag, but \hcn\ and \hcop\ trace it better for higher \av, confirming the view that dense line tracers (\hcn\ and \hcop) are more sensitive to the high column density regions than is \coo.  
    
\end{enumerate}

\begin{acknowledgments}
The authors thank the referee for constructive comments which improved the clarity of the paper. We thank the staff of the TRAO for support during the course of these observations. SP thanks DST-INSPIRE Fellowship (No. IF180092) of the Department of Science and Technology to support the Ph.D. program. NJE thanks the Department 
of Astronomy at the University of Texas at Austin for ongoing research support. We are grateful to Adam Ginsburg for sharing information on BGPS.
\end{acknowledgments}

\vspace{5mm}
\facilities{TRAO 14-m Telescope, FCRAO, BGPS, \textit{Herschel}}

\software{astropy \citep{2013A&A...558A..33A,2018AJ....156..123A,2022arXiv220614220T},
          numpy \citep{2011CSE....13b..22V},
          matplotlib \citep{2007CSE.....9...90H},
          GILDAS \citep{2018ssdd.confE..11P},
          DS9 \citep{2003ASPC..295..489J, 2000ascl.soft03002S}}

\begin{deluxetable*}{l c c l l c c c c c c}[h]
    \tablenum{1}
    \tabletypesize{\footnotesize}
    \tablecaption{Sample of Clouds \label{tab:cloud details}}
    \tablewidth{0pt}
    \tablehead{
    \colhead{Target} & \colhead{$l$} & \colhead{$b$} &  \colhead{\rg} & \colhead{$D$} & \nocolhead{size} & \colhead{\mcloud}  & \colhead{SpT} & \colhead{\mbox{12+log(O/H)}} & \colhead{Map Size}    & \colhead{\av\tablenotemark{*}}\\
    \colhead{} & \colhead{(deg)} & \colhead{(deg)} & \colhead{(kpc)} & \colhead{(kpc)} & \nocolhead{(pc)} & \colhead{(\eten4\ \msun)} & \colhead{(Ref)} & \colhead{(Ref)} & \colhead{($\ \ '\ \times \ \ '$)}  & \colhead{(mag)}
    }
    \startdata
        \clouda\tablenotemark{a} & $150.5886$ & $-0.8570$ & $10.88^{+0.25}_{-0.26}$ & $2.96^{+0.17}_{-0.15}$ &  & $18.5$  & O4V (1)   & 8.37 (8) &  15 $\times$ 15    & 1.0 \\
        \cloudb\tablenotemark{a} &  $151.3098$ & $1.9045$ & $11.87^{+0.37}_{-0.34}$ & $4.02^{+0.27}_{-0.25}$ &    & $6.8$   & B0V (2) & 8.43 (8) & 10 $\times$ 20      & 0.3 \\
        \cloudc\tablenotemark{b} &  $155.3375$ & $2.6345$ & $14.76\pm1.30$ & $6.65^{+1.36}_{-1.26}$ &    & $19.0$   & O7 (3)  & 8.34 (9) & 13 $\times$ 14 & 0.5 \\
        \cloudd\tablenotemark{c} &  $169.1432$ & $-1.0475$ & $10.72\pm0.19$ & $2.56\pm0.09$  &    & $11.6$    & O8V (4) & 8.39 (8) & 11 $\times$ 33  & 0.4 \\
        \cloude\tablenotemark{b} &  $173.6682$ & $2.7799$ & $9.85^{+0.16}_{-0.17}$  & $1.66\pm0.07$ &    & 41.1       & O9.5V (5)    & 8.54 (8) & 30 $\times$ 30  & 1.5 \\
        \cloudf\tablenotemark{c} &  $189.8418$ & $0.3156$ & $10.09^{+0.12}_{-0.11}$ & $1.92^{+0.12}_{-0.11}$  &    & 49.0     & O6.5V,O9.5V,B V (6)   & 8.51 (8) & 30 $\times$ 30 & 1.5 \\
        \cloudg\tablenotemark{c} &  $192.7935$ & $0.0290$ & $10.13\pm0.21$ & $1.96^{+0.12}_{-0.09}$  &    & 51.5       &O9.5V,B0.5V (7)           & 8.69 (8) & 50 $\times$ 34 & 1.5 \\
    \enddata
    \tablecomments{(1)\citet{2016ApJS..224....4M}, (2)\citet{moffat1979rotation}, (3)\citet{2014ApJS..211...10S}, (4)\citet{chini1984galactic}, (5)\citet{georgelin1973observations}, (6)\citet{2012MNRAS.424.2486J}, (7)\citet{2011ApJ...738..156O}, (8)\cite{2018PASP..130k4301W}, (9)\cite{2022MNRAS.tmp...13M}}
    \textbf{Data availability:} \tablenotetext{a}{\cotw, \coo} \tablenotetext{b}{\cotw, \coo, BGPS} \tablenotetext{c}{\cotw, \coo, BGPS, Herschel}
    \tablenotetext{*}{\av \ values corresponding to the threshold value of 5 times the RMS noise in the \coo\ map.}
\end{deluxetable*}

\begin{deluxetable*}{l c c c c c c c c c c}[h]
    \tablenum{2}
    \tabletypesize{\footnotesize}
    \tablecaption{Line Luminosities versus $\av \ge 8 \ \mbox{mag}$ \label{tab:av8}}
    \tablewidth{0pt}
    \tablehead{
    \colhead{Source} & \colhead{$N_{\rm{in}}/N_{\rm tot}$} & \colhead{Log $L_{\rm tot}$} & \colhead{Log $L_{\rm in}$} & \colhead{\fl} & \colhead{Log $L_{\rm tot}$} & \colhead{Log $L_{\rm in}$} & \colhead{\fl} & \colhead{Log $L_{\rm tot}$} & \colhead{Log $L_{\rm in}$} & \colhead{\fl} \\
    \colhead{} & \colhead{} & \colhead{\hcn} & \colhead{\hcn} & \colhead{\hcn} & \colhead{\hcop} & \colhead{\hcop} & \colhead{\hcop} & \colhead{\coo} & \colhead{\coo} & \colhead{\coo} 
    }
    \startdata
        \clouda & 0.067 & $1.88^{+0.05}_{-0.05}$ & $1.14^{+0.05}_{-0.05}$ & 0.180(0.008) & $1.84^{+0.05}_{-0.05}$ & $1.05^{+0.05}_{-0.05}$  & 0.164(0.006) & $2.54^{+0.05}_{-0.05}$ & $1.76^{+0.05}_{-0.05}$ & 0.167(0.004)\\
        \cloudb & 0.061 & $1.31^{+0.06}_{-0.06}$ & $0.63^{+0.06}_{-0.06}$ & 0.209(0.011) & $1.39^{+0.05}_{-0.06}$ & $0.62^{+0.05}_{-0.06}$ & 0.170(0.004) & $2.04^{+0.05}_{-0.06}$ & $1.18^{+0.07}_{-0.07}$ & 0.138(0.014)\\
        \cloudc & \nodata & $1.89^{+0.15}_{-0.21}$ & \nodata              & \nodata        & $1.95^{+0.15}_{-0.21}$ & \nodata                & \nodata & $2.49^{+0.15}_{-0.21}$ & \nodata & \nodata \\
        \cloudd & 0.008 & $1.40^{+0.05}_{-0.05}$ & $-0.44^{+0.08}_{-0.09}$ & 0.014(0.003) & $1.29^{+0.04}_{-0.05}$ & $-0.75^{+0.10}_{-0.13}$ & 0.009(0.002) & $2.77^{+0.03}_{-0.03}$ & $1.02^{+0.07}_{-0.08}$ & 0.018(0.003) \\
        \cloude & 0.164 & $2.45^{+0.04}_{-0.04}$ & $2.18^{+0.03}_{-0.04}$ & 0.537(0.011) & $2.43^{+0.03}_{-0.04}$ & $2.10^{+0.03}_{-0.04}$ & 0.471(0.007) & $3.09^{+0.03}_{-0.04}$ & $2.64^{+0.03}_{-0.04}$ & 0.355(0.002) \\
        \cloudf & 0.265 & $2.53^{+0.05}_{-0.05}$ & $2.35^{+0.05}_{-0.05}$ & 0.661(0.025) & $2.48^{+0.05}_{-0.05}$ & $2.27^{+0.05}_{-0.05}$ & 0.608(0.010) & $3.32^{+0.05}_{-0.05}$ & $3.08^{+0.05}_{-0.05}$ & 0.570(0.003)\\
        \cloudg & 0.107 & $2.50^{+0.05}_{-0.04}$ & $2.16^{+0.05}_{-0.04}$ & 0.459(0.008) & $2.49^{+0.05}_{-0.04}$ & $2.06^{+0.05}_{-0.04}$ & 0.371(0.006) & $3.21^{+0.05}_{-0.04}$ & $2.64^{+0.05}_{-0.04}$ & 0.269(0.002)\\
        \tableline
        Mean     & 0.112   & 1.99    &  1.34    &	0.343   &	1.98    &	1.22    &   0.299   &	2.78    &	2.05	&   0.253\\
        Std Dev. & 0.083   & 0.48    &	1.01	&   0.225   &	0.47    &	1.07	&   0.204   &	0.42	&   0.78	&   0.177\\
        Median	 & 0.087   & 1.89    &	1.65    &	0.334   &	1.95    &	1.56	&   0.271   &	2.77	&   2.20	&   0.218\\
    \enddata
    \tablecomments{1. Units of luminosities are \kkms pc$^2$.}
\end{deluxetable*}

\begin{deluxetable*}{l c c c c c c c c c c}[h]
    \tablenum{3}
    \tabletypesize{\footnotesize}
    \tablecaption{Line Luminosities versus BGPS Emission }\label{tab:bgps}
    \tablewidth{0pt}
    \tablehead{
    \colhead{Source} & \colhead{$N_{\rm{in}}/N_{\rm tot}$} & \colhead{Log $L_{\rm tot}$} & \colhead{Log $L_{\rm in}$} & \colhead{\fl} & \colhead{Log $L_{\rm tot}$} & \colhead{Log $L_{\rm in}$} & \colhead{\fl} & \colhead{Log $L_{\rm tot}$} & \colhead{Log $L_{\rm in}$} & \colhead{\fl} \\
    \colhead{} & \colhead{} & \colhead{\hcn} & \colhead{\hcn} & \colhead{\hcn} & \colhead{\hcop} & \colhead{\hcop} & \colhead{\hcop} & \colhead{\coo} & \colhead{\coo} & \colhead{\coo} 
    }
    \startdata
        \cloudc & 0.352 & $1.89^{+0.15}_{-0.21}$ & $1.75^{+0.15}_{-0.21}$ & 0.728(0.045) & $1.95^{+0.15}_{-0.21}$ & $1.80^{+0.15}_{-0.21}$ & 0.717(0.025) & $2.49^{+0.15}_{-0.21}$ & $2.06^{+0.15}_{-0.21}$ & 0.376(0.017) \\
        \cloudd & 0.049 & $1.40^{+0.05}_{-0.05}$ & $0.84^{+0.03}_{-0.03}$ & 0.277(0.027) & $1.29^{+0.04}_{-0.05}$ & $0.89^{+0.03}_{-0.03}$ & 0.391(0.034) & $2.77^{+0.03}_{-0.03}$ & $1.60^{+0.04}_{-0.04}$ & 0.068(0.004) \\
        \cloude & 0.118 & $2.45^{+0.04}_{-0.04}$ & $2.08^{+0.03}_{-0.04}$ & 0.433(0.009) & $2.43^{+0.03}_{-0.04}$ & $2.02^{+0.03}_{-0.04}$ & 0.386(0.006) & $3.09^{+0.03}_{-0.04}$ & $2.49^{+0.03}_{-0.04}$ & 0.248(0.001) \\
        \cloudf & 0.165 & $2.53^{+0.05}_{-0.05}$ & $2.26^{+0.05}_{-0.05}$ & 0.533(0.014) & $2.48^{+0.05}_{-0.05}$ & $2.18^{+0.05}_{-0.05}$ & 0.496(0.008) & $3.32^{+0.05}_{-0.05}$ & $2.90^{+0.05}_{-0.05}$ & 0.381(0.002) \\
        \cloudg & 0.085 & $2.50^{+0.05}_{-0.04}$ & $2.12^{+0.05}_{-0.04}$ & 0.418(0.007) & $2.49^{+0.05}_{-0.04}$ & $2.01^{+0.05}_{-0.04}$ & 0.331(0.005) & $3.21^{+0.05}_{-0.04}$ & $2.53^{+0.05}_{-0.04}$ & 0.208(0.001)\\
        \tableline
        Mean	&   0.154    &	2.15 &	1.81     &  0.478	& 2.13	& 1.78	& 0.464	& 2.97	& 2.32	& 0.256 \\
        Std Dev.&  	0.106	&   0.44 &	0.51	 &  0.149	& 0.46	& 0.46	& 0.137	& 0.31	& 0.50	& 0.116 \\
        Median	&   0.118	&   2.45 &	2.08     &	0.433	& 2.43	& 2.01	& 0.391	& 3.09	& 2.49	& 0.248 \\
    \enddata
    \tablecomments{1. Units of luminosities are \kkms pc$^2$.}
\end{deluxetable*}

\begin{deluxetable*}{l c c c c c c c c c c}[h]
    \tablenum{4}
    \tabletypesize{\footnotesize}
    \tablecaption{\mdense\ versus Luminosities \label{tab:Mdense}}
    \tablewidth{0pt}
    \tablehead{
    \colhead{Source} & \colhead{$\mathrm{Log\ \mdense}$} & \colhead{Log $L_{\rm tot}$} & \colhead{Log $L_{\rm in}$} & \colhead{$\mathrm{Log \ \alpha_{tot}}$} & \colhead{$\mathrm{Log \ \alpha_{in}}$} & \colhead{Log $L_{\rm tot}$} & \colhead{Log $L_{\rm in}$} & \colhead{$\mathrm{Log \ \alpha_{tot}}$} & \colhead{$\mathrm{Log \ \alpha_{in}}$} \\
    \colhead{} & \colhead{\msun} & \colhead{\hcn} & \colhead{\hcn} & \colhead{\hcn} & \colhead{\hcn} & \colhead{\hcop} & \colhead{\hcop} & \colhead{\hcop} & \colhead{\hcop} 
    }
    \startdata
        \cloudc &  $3.06^{+0.16}_{-0.25}$   & $1.89^{+0.15}_{-0.21}$ & $1.75^{+0.15}_{-0.21}$ & $1.17\pm 0.09$ & $1.31\pm 0.08$ & $1.95^{+0.15}_{-0.21}$ & $1.80^{+0.15}_{-0.21}$ & $1.11\pm0.08$  & $1.26\pm0.08$\\
        \cloudd &  $2.19^{+0.07}_{-0.08}$   & $1.40^{+0.05}_{-0.05}$ & $0.84^{+0.03}_{-0.03}$ & $0.80\pm 0.07$ & $1.35\pm 0.06$ & $1.29^{+0.04}_{-0.05}$ & $0.89^{+0.03}_{-0.03}$ & $0.90\pm0.07$ & $1.31\pm0.06$\\
        \cloude &  $3.22^{+0.05}_{-0.05}$   & $2.45^{+0.04}_{-0.04}$ & $2.08^{+0.03}_{-0.04}$ & $0.77\pm 0.04$ & $1.13\pm 0.03$ & $2.43^{+0.03}_{-0.04}$ & $2.02^{+0.03}_{-0.04}$ & $0.79\pm0.04$ & $1.20\pm0.03$\\
        \cloudf &  $3.53^{+0.06}_{-0.07}$   & $2.53^{+0.05}_{-0.05}$ & $2.26^{+0.05}_{-0.05}$ & $1.00\pm 0.04$ & $1.28\pm 0.04$ & $2.48^{+0.05}_{-0.05}$ & $2.18^{+0.05}_{-0.05}$ & $1.05\pm0.04$ & $1.35\pm0.04$\\
        \cloudg &  $3.43^{+0.06}_{-0.06}$   & $2.50^{+0.05}_{-0.04}$ & $2.12^{+0.05}_{-0.04}$ & $0.93\pm 0.04$ & $1.31\pm 0.04$ & $2.49^{+0.05}_{-0.04}$ & $2.01^{+0.05}_{-0.04}$ & $0.93\pm0.04$ & $1.41\pm0.04$\\
        \tableline
        Mean	&  3.09     &	2.15 &	1.81     & 0.93 & 1.28	& 2.13	&1.78	& 0.96  &   1.31 \\
        Std Dev.&  0.53	    &   0.44 &	0.51	 & 0.16 & 0.08	& 0.46	&0.46	& 0.13  &   0.08\\
        Median	&  3.22 	&   2.45 &	2.08     & 0.93	& 1.31	& 2.43	&2.01	& 0.93  &   1.31\\
    \enddata
    \tablecomments{1. Units of luminosities are \kkms pc$^2$.\\
    2. Units of conversion factors are $\msun \ (\rm K\ km\ s^{-1}\ pc^{2})^{-1}$}
\end{deluxetable*}


\clearpage
\bibliographystyle{aasjournal}
\bibliography{ms}{}

\appendix

\section{Target Details} \label{app:target}
All of the sources were identified in the second catalog of visible \hii\ regions by
\citet{1959ApJS....4..257S}.
\subsection{Sh2-206}
 Sh2-206 (hereafter \clouda) was mapped with the center position $l= 150.5886$, $b=-0.8570$ with a map size $15'\times 15'$. This \hii\ region is not excited by a cluster, but by a single massive star BD$+50886$ \citep{georgelin1973observations}. 
 An ionization front forms the bright central region of this target and it surrounds the O4V exciting star to the south and west  \citep{2000MNRAS.311..329D}. 
 The heliocentric distance of this region is $2.96^{+0.17}_{-0.15}$ kpc, measured from the parallax information of the massive star and the Galactocentric distance is $\rg=10.88^{+0.25}_{-0.26}$ kpc \citep{2022MNRAS.tmp...13M}.

 We use the velocity range -27 to -19 \kms\ to produce the column density map from \coo\ and integrated intensity map for \hcop. For \hcn\ we use the velocity range -34 to -14 \kms\ for \hcn\ integrated intensity map. The presence of a dense region ($\av \ge 8$ mag) is indicated with the white contour on top of the integrated intensity of \hcn\ and \hcop in Figure \ref{fig:panel}(\clouda). There is strong detection of \hcn\ and \hcop\ emission in both \textbf{in} and \textbf{out} regions for extinction in the spectrum plots (Figure \ref{fig:spectrum}a). The \fl\ value based on \av\ analysis is higher for \hcn, followed by \coo\ and \hcop. This target lacks data in millimeter-wave continuum and Herschel. 

\subsection{Sh2-208}
 Sh2-208 (hereafter \cloudb) was mapped with center position $l=151.3098, \ b=1.9045$ with a map size $10' \times 20'$. This \hii\ region is located at a heliocentric distance of $4.02_{+0.27}^{-0.25}$ kpc in an interarm island between the Cygnus and Perseus arms \citep{2016AJ....151..115Y} and the corresponding Galactocentric distance is $11.87_{+0.37}^{-0.34}$ kpc. This HII region is located in a sequential star forming region and the probable dominant star is GSC 03719-00517 \citep{2016AJ....151..115Y}.
 
 We use the velocity range -32 to -28 \kms\ to produce the column density map from \coo\ and integrated intensity map for \hcop. For \hcn\ we use the velocity range -40 to -23 \kms\ for integrated intensity map. Figure \ref{fig:panel} shows the integrated intensity plot of \hcn\ and \hcop, and the white contours show the area above $\av=8$ mag. The extinction criterion of dense gas is in good agreement with the emission of \hcn\ and \hcop. Figure \ref{fig:panel}(b) shows strong emission is coming from the \textbf{in} region  for both the lines. 
\subsection{Sh2-212}
 Sh2-212 (hereafter \cloudc) is a bright optically-visible HII region, ionized by a cluster (NGC 1624) containing an O7 type star. 
 \cloudc\ appears circular in the optical band around the central cluster. The molecular cloud appears as a semi-circular shape in the southern part of the region. The region contains multiple substructures and indicates that it is evolving in a non-homogeneous medium \citep{2008A&A...482..585D}. This region is a likely example of the collect-and-collapse process triggering massive star formation (see \citealt{2008A&A...482..585D, 2011MNRAS.411.2530J} for details).
 
 Our map covered a region of $13'\times14'$ centered on $l=155.3375$, $b=2.6345$.  This target lies $\sim300$ pc above the Galactic plane. The heliocentric distance is $6.65_{+1.36}^{-1.26}$ kpc \citep{2022MNRAS.tmp...13M}. This target has the highest Galactocentric distance ($14.76 \pm 1.30$ kpc) in the sample and lowest metallicity.
 
 To derive the column density map from \coo\ and for the integrated intensity map of \hcop, we use the velocity integration range -39 to -30.5 \kms. For \hcn\ we use the velocity range -44 to -26 \kms\ for integrated intensity map. Figure \ref{fig:panel} (\cloudc) shows the integrated intensity plots of \hcn\ and \hcop. For this target there is no region with $\av \ge 8$ mag, so there is no white contour present in the figure. But the position of BGPS mask (shown in black contours) and \hcn, \hcop\ shows good agreement. Figure \ref{fig:spectrum}(c) shows the spectra for both extinction and BGPS based analysis. 

\subsection{Sh2-228}
 Sh2-228 (hereafter \cloudd) was mapped in Galactic co-ordinate with the center position $l= 169.1432$, $b=-1.0475$ with a map size $11'\times 33'$. This \hii\ region is located on the Perseus arm near the large Auriga OB1 association. It is a vast region in which star formation is active. A young cluster IRAS 05100+3723 is associated with the \hii\ region \cloudd\ \citep{2022ApJ...926...16Y}. The heliocentric distance of this target is $2.56\pm 0.09$ kpc and the corresponding Galactocentric distance $\rg=10.72\pm0.19$ kpc.  
 
 To derive the column density map from \coo\ and for the integrated intensity map of \hcop, we use the velocity integration range -12 to -3.5 \kms. For \hcn\ we use the velocity range -16 to 2 \kms for integrated intensity map. Figure \ref{fig:panel}(\cloudd) shows the position of $\av \ge 8$ mag region and BGPS clumps positions with white and black contours respectively on the integrated intensity maps of \hcn\ and \hcop. The emission of \hcn\ and \hcop\ show good agreement with the BGPS position, but there is no strong emission from \hcn, \hcop\ in the $\av \ge 8$ mag region (see spectra in Figure \ref{fig:spectrum}(d) ). 
 The lines of \coo\ in the region of the BGPS peak are slightly below the value corresponding to $\av = 8$, instead indicating $\av = 6$ or 7 mag. There are two peaks in the Herschel column density maps, one at the position of BGPS clump and the other  below the region where \coo\ peaks. The lack of strong \coo\ emission at a peak of BGPS, \hcn, and \hcop\ is unusual. We examined the spectra of \coo\ at the BGPS peak for evidence of self-absorption, but none was obvious.  
 The \fl\ value based on BGPS analysis is  higher for \hcn, followed by \hcop\ and \coo. The line-to-mass ratio versus dust-based extinction plot (Figure \ref{fig:herschel}a) shows that \hcop\ and \hcn\ are better tracers in the dense region than \coo.   
 \subsection{Sh2-235}
 Sh2-235 (hereafter \cloude) is an extended \hii\ region with active star formation and it is located in the Perseus spiral arm \citep{1996ApJ...463..630H}. 
 This star forming region consists of ``S235main" and S235A, S235B, and S235C regions (hereafter ``S235ABC”) \citep{2011MNRAS.414.1526D}.  A single massive star BD$+351201$ of O9.5V type is ionizing the ``S235main", while the compact \hii\ regions in ``S235ABC" regions are excited by B1V–B0.5V stars (\citet{2014MNRAS.439.3719C, 2017ApJ...849...65D} and references therein).
 This cloud has two velocity components, ``S235main'' with velocity range $-24$ to $-18$ \kms\ and ``S235ABC" with velocity range $-18$ to $-13$ \kms. These two clouds are interconnected by a bridge feature with less intensity \citep{1981ApJ...246..394E, 2017ApJ...849...65D}.  This region is also known for the star formation activities influenced by the cloud-cloud collision  mechanism \citep{2017ApJ...849...65D}.
 
  We mapped $30'\times 30'$ region with the center $l=173.6682$, $b=2.7799$. The heliocentric distance is $1.66\pm0.07$ kpc, derived from the parallax information of the ionizing star BD$+351201$ from Gaia EDR3 data. The corresponding Galactocentric distance  $\rg = 9.85_{+0.16}^{-0.17}$ kpc \citep{2022MNRAS.tmp...13M}. 

 To derive the column density map from \coo\ and for the integrated intensity map of \hcop, we use the velocity integration range -24.5 to -12 \kms. For \hcn\ we use the velocity range -32 to -7 \kms\ for integrated intensity map. 
 The 2 velocity component at -20 \kms\ and -17.3 \kms\ are not separable at some positions, so the integration range we consider covers both the components. 
 All the tracers (\hcn, \hcop, extinction and BGPS emission) agree with each other (Figure \ref{fig:panel}(\cloude)). The white contour ($\av \ge 8$ mag) covers larger area than the BGPS region (black contours). 
 There are strong detections of \hcn\ and \hcop\ emission in both \textbf{in} and \textbf{out} regions for extinction and BGPS in the spectrum plots (Figure \ref{fig:spectrum}e).
 
 \subsection{Sh2-252}
 Sh2-252 (hereafter \cloudf), also known as NGC 2175 is an optically bright, evolved \hii\ region with size $\sim20$ pc. 
 This region is part of Gemini OB1 association. The central star HD 42088 of spectral type O6.5 V is the main ionizing source \citep{2012MNRAS.424.2486J}.
 This region also contains four compact \hii\ (C\hii) regions (A, B, C and E) \citep{felli1977aperture} and powered by ionizing sources with spectral type later than O6.5V in each of them. This cloud has several regions of recent star formation activity.
 
 Table 4 of \citet{2012MNRAS.424.2486J} listed the massive stars in each of these clumps with their spectral types. We take the parallax information of those stars (ID 1-6 and 11-12 from the Table 4 of \citealt{2012MNRAS.424.2486J})  from Gaia EDR3 and calculate the distance by averaging all the parallaxes. 
 The distance is $1.92^{+0.12}_{-0.11}$ kpc and the corresponding Galactocentric distance $\rg=10.09_{+0.12}^{-0.11}$ kpc. This source was mapped with center position $l=189.8418$, $b=0.3156$ with a map size $30'\times30'$.  

 We use the velocity range 2 to 14 \kms\ to produce the column density map from \coo\ and integrated intensity map of \hcop. For \hcn\ we use the velocity range -4 to 20 \kms\ for integrated intensity map. Figure\ref{fig:panel}(\cloudf) shows the contour of $\av \ge 8$ mag (in white) and BGPS clumps (in black) in the integrated intensity maps of \hcn, \hcop\ and all the tracers are in good agreement. 
 There is strong detection of \hcn\ and \hcop\ emission in both \textbf{in} and \textbf{out} regions for extinction and BGPS in the spectrum plots (Figure \ref{fig:spectrum}f).

 \subsection{Sh2-254 Complex}
 This region is a complex with five \hii\ regions (Sh2-254, Sh2-255, Sh2-256, Sh2-257 and Sh2-258) and part of Gemini OB association. This is a sequential star-forming region triggered by expanding \hii\ regions \citep{2007IAUS..237..396B, 2008ApJ...682..445C, 2011ApJ...738..156O}. These five evolved \hii\ regions are projected on a 20 pc long filament. The exciting stars present in S254, S255, S256, S257 and S258 have spectral types of O9.6 V, B0.0 V, B0.9 V, B0.5V and B1.5V \cite{2008ApJ...682..445C}. \citet{2021MNRAS.506.4447L} studied the link between the gas-dust constituents and the young stellar objects (YSOs) distribution. 
 They identified high-density gas (\hcop(\jj10) and CS(\jj21)) in the inter-clump bridge position between the star clusters S255N and S256-south and suggested that the clusters have an evolutionary link.
 We mapped $50'\times34'$ area of Sh2-254 complex with the center position $l=192.7935$, $b=0.0290$. The Galactocentric distance is $\rg=10.13\pm0.21$ kpc and the heliocentric distance is $1.96_{+0.12}^{-0.09}$ kpc based on GAIA EDR3 \citep{2022MNRAS.tmp...13M}.
 
 We use the velocity range 4.5 to 11.5 \kms\ to produce the column density map from \coo\ and integrated intensity map of \hcop. For \hcn\ we use the velocity range -4 to 17 \kms\ for integrated intensity map. The contour plot (Figure \ref{fig:panel}(\cloudg)) for this cloud shows  good agreement among \hcn, \hcop, extinction and BGPS. 
 There are strong detections of \hcn\ and \hcop\ emission in both \textbf{in} and \textbf{out} regions for extinction and BGPS in the spectrum plots (Figure \ref{fig:spectrum}g).

\section{Reduction Details} \label{app:reduction}
Our standard procedure of data reduction is as follows. First we check the data to find the velocity intervals with significant emission ($v_{win}$). Then we exclude the window of velocity having emission while removing a baseline order, also using only enough velocity range to get good baseline on each end ($v_{sp}$). After that we made spectral cubes in FITS format for further analysis. The values of the total velocity range and excluded windows are shown in Table \ref{lineprops}. 
We experimented on the velocity range ($v_{sp}$) and baseline order to get the best fit.

\paragraph{\textbf{Change in baseline order}}
First we have experimented on the baseline order removal. For a same target, by keeping $v_{win}$ and $v_{sp}$ same, we tried both first and second order baselines. For most of the cases, the second order baseline removal was superior. 
\paragraph{\textbf{Change in $v_{sp}$ range}}
Further we change the velocity range of $v_{sp}$ (while keeping the baseline removal order fixed) for each target with 2 different set - \textit{(i) wider range of $v_{sp}$} and \textit{(ii) narrow range of $v_{sp}$} to see the effect on final results.

 In the final analysis, we used baselines of order 2 with wider range of $v_{sp}$.
 Table \ref{tab:reduction} shows the choice of velocity ranges and lists the positions of all the peaks in terms of offset from the center of the map, with offsets in Galactic coordinates.
\begin{longtable}{c c c c c c c c }
\caption{Data Reduction Details} \label{tab:reduction}\\
\hline
\hline
Source      &       Line        & $v_{sp}$      &       $v_{win}$             & Peak Offset             & Notes \\
            &                   & (\kms)        &       (\kms)                & (arcsec)                &    \\
\hline
\endfirsthead
\multicolumn{8}{c}%
{\tablename\ \thetable\ -- \textit{Continued from previous page}} \\
\hline
\hline
Source      &       Line        & $v_{sp}$      &       $v_{win}$             & Peak Offset               & Notes \\
            &                   & (\kms)        &       (\kms)                & (arcsec)                  &    \\
\hline
\endhead
\hline \multicolumn{8}{r}{\textit{Continued on next page}} \\
\endfoot
\hline
\endlastfoot
       
    \clouda & \hcn &  -45, 0   & -32, -14   &  280, 280 &            Peak position 1 \\
    \clouda & \hcop & -45, 0  & -27, -19    &  280, 280 &           Peak position 1\\
    \clouda & \hcn & -45, 0    & -32, -14    &  180, 120 &             Peak position 2 \\
    \clouda & \hcop &  -45, 0 & -27, -19    &  180, 120 &           Peak position 2\\
    \clouda & \hcn & -45, 0    & -32, -14    &  -40, -280&             Peak position 3 \\
    \clouda & \hcop &  -45, 0 & -27, -19    &  -40, -280&           Peak position 3\\
    \\
    \cloudb & \hcn & -55, -5    & -40, -23         & -100, 200           &    \\
    \cloudb & \hcop & -55, -5  & -34, -27         & -100, 200           &    \\
    \\
    \cloudc & \hcn & -60, -10   & -44, -26         & 0, -160            &    Peak position 1 \\
    \cloudc & \hcop & -60, -10 & -40, -29         & 0, -160            &  Peak position 1 \\
    \cloudc & \hcn & -60, -10   & -44, -26         & 120, -40           &    Peak position 2 \\
    \cloudc & \hcop & -60, -10 & -40, -29         & 120, -40           &  Peak position 2 \\
    \\
    \cloudd & \hcn & -30, 20    & -16, 3          & 160, 520           &    \\
    \cloudd & \hcop & -30, 20  & -11, -2         & 160, 520            &    \\
    \\
    \cloude & \hcn & -50, 10    & -32, -7         & -160, 100             &  Peak position 1  \\
    \cloude & \hcop & -50, 10  & -26, -12         & -160, 100             & Peak position 1    \\
    \cloude & \hcn &  -50, 10   & -32, -7          & 180, -300             &  Peak position 2  \\
    \cloude & \hcop & -50, 10  & -26, -12         & 180, -300             & Peak position 2    \\
    \cloude & \hcn &  -50, 10   & -32, -7          &   0, 340              &  Peak position 3  \\
    \cloude & \hcop & -50, 10  & -26, -12         &   0, 340              & Peak position 3    \\
    \cloude & \hcn &  -50, 10   & -32, -7          & -180, 360             &  Peak position 4 \\
    \cloude & \hcop &  -50, 10 & -26, -12         & -180, 360             & Peak position 4    \\
    \\    
    \cloudf & \hcn &  -20, 40   & -4, 18           & -220, 100            & Peak position 1   \\
    \cloudf & \hcop & -20, 40  & 2, 14            & -220, 100            & Peak position 1   \\
    \cloudf & \hcn &  -20, 40   & -4, 20           & -120, 120            & Peak position 2   \\
    \cloudf & \hcop & -20, 40  & 0, 15            & -120, 120            & Peak position 2   \\
    \cloudf & \hcn &  -20, 40   & -4, 20           &  400, 60             & Peak position 3   \\
    \cloudf & \hcop & -20, 40  & 0, 15            &  400, 60             & Peak position 3   \\
    \cloudf & \hcn &  -20, 40   & -4, 20           &  0, 640              & Peak position 4   \\
    \cloudf & \hcop & -20, 40  & 0, 15            &  0, 640              & Peak position 4   \\
    \cloudf & \hcn &  -20, 40   & -4, 20           &  400, -300           & Peak position 5   \\
    \cloudf & \hcop & -20, 40  & 0, 15            &  400, -300           & Peak position 5   \\
    \cloudf & \hcn &  -20, 40   & -4, 20           & -620, -520           & Peak position 6   \\
    \cloudf & \hcop & -20, 40  & 0, 15            & -620, -520           & Peak position 6   \\
    \\    
    \cloudg & \hcn   & -25, 35   & -4, 17           & -720, -280            &  Peak position 1  \\
    \cloudg & \hcop & -25, 35   &  2, 14           & -720, -280            & Peak position 1    \\
    \cloudg & \hcn   & -25, 35   & -4, 17           &  740, 360             &  Peak position 2  \\
    \cloudg & \hcop & -25, 35   &  2, 14           &  740, 360             & Peak position 2    \\
    \cloudg & \hcn   & -25, 35   & -4, 17           & -120, 240             &  Peak position 3  \\
    \cloudg & \hcop & -25, 35   &  2, 14           & -120, 240             & Peak position 3    \\
    \cloudg & \hcn   & -25, 35   & -4, 17           &  220, 280             &  Peak position 4 \\
    \cloudg & \hcop & -25, 35   &  2, 14           &  220, 280             & Peak position 4    \\
\label{lineprops}
\end{longtable}

\section{The \hh\ column density map derived from \cotw\ and \coo} \label{app:H2}
 
 To measure the molecular hydrogen (\hh) column density map from \cotw\ and \coo, we assume that both the molecular lines  arise from the same region of the clouds. We follow the steps from \citet{2013MNRAS.431.1296R} and \citet{2010ApJ...721..686P}. \cotw\ line is optically thick and its isotopologue \coo\ is optically thin. We assume that they both have same excitation temperature ($T_{ex}$). So we derive the $T_{ex}$ using the peak brightness temperature ($T_{12,pk}$) of the optically thick \cotw\ line. 
 \begin{equation}
    T_{ex} = \frac{5.53}{\mathrm{ln}\left (1+ \frac{5.53}{T_{12,pk}+0.83}  \right )}
 \end{equation}

 We derive the optical depth of \coo\ in the upper rotational state $(J=1)$, $(\tau_{13,1})$ using the main-beam brightness temperature of \coo, $(T_{13})$ and substitute $T_{ex}$ derived from the \cotw\ observations into the following equation 
\begin{equation}
    \tau_{13,1}(v) = - \mathrm{ln} \Bigg[1-\frac{T_{13}(v)}{5.29} \bigg( \bigg[\mathrm{exp(5.29/T_{ex})-1} \bigg ]^{-1} -0.16 \bigg )^{-1} \Bigg ]
\end{equation}

 The optical depth in \coo\ is then converted into a upper level column density as follows
 \begin{equation}
    N_{U} = \frac{8\pi k \nu^{2}}{hc^{3}A_{UL}C(T_{ex})}\Bigg[\frac{\int \tau_{13,1}(v)dv}{\int(1-\mathrm{exp}(-\tau_{13,1}(v)))dv} \Bigg]\int T_{13}(v)dv
 \end{equation}
 where $A_{UL}$ is the Eienstein A coefficient and $\nu$ is the frequency of the \jj10 transition. Here we incorporated the correction factor $C(T_{ex})= \left (1- \frac{e^{T_0/T_{ex}}-1}{e^{T_0/T_{bg}}-1} \right)$, which is equal to unity in the limit $T_{bg}\to 0$.

 The column density of the upper level $(J=1)$ is related to the  total \coo\ column density by the following equation
 \begin{equation}
    N_{\coo}= N_{U} \frac{Z}{2J+1}\mathrm{exp}[h B_{0} J(J+1)/kT_{ex}]
 \end{equation}
 To calculate the upper level column density, we use the rotational constant for \coo\ $B_{0}=5.51\times10^{10}$ s$^{-1}$ and use the partition function $Z$ which is defined as

 \begin{equation}
    Z= \sum_{J=0}^{\infty}(2J+1)\mathrm{exp}[-hB_{0}J(J+1)/kT_{ex}]
 \end{equation}
This equation using a single $T_{ex}$ assumes that all levels have the same $T_{ex}$ (the so-called LTE approximation).

The assumption that $T_{ex}$ is the same for \coo\ and \cotw\ (in equation C1) can fail. If the excitation temperature of \coo\ is higher (lower) than that of \cotw, the optical depth of \coo\ will be over-estimated (under-estimated) when the $T_{ex}$ from \cotw\ is used in equation C2. For clouds heated from the outside by hot stars, the front of the cloud is likely to be hotter than the inside, leading to an underestimate of the optical depth of \coo.

\citet{2021A&A...645A..26R} checked the validity of the simple assumption of equal $T_{ex}$ in the Orion B molecular cloud by using Cramer Rao Bound (CRB) technique to estimate different physical properties (column density, excitation temperature, velocity dispersion etc.) in the framework of LTE radiative transfer model. They found that isotopologues had different excitation temperatures;  \cotw\ (30 K) had higher $T_{ex}$ than \coo\ (17 K), so that $T_{ex}(\cotw)/T_{ex}(\coo) \sim 1.7$. If a similar situation applies to the more opaque parts of our clouds, \coo\ is likely to underestimate the column density.

\end{document}